\documentclass[
reprint,
groupedaddress,
amsmath,amssymb,
prmaterials,
floatfix,
superscriptaddress,
]{revtex4-2}

\usepackage{amsmath}
\usepackage{bm} 
\usepackage{amsfonts}
\usepackage{graphicx}
\usepackage{amssymb}
\usepackage[caption=false]{subfig}
\usepackage{hyperref}
\usepackage{cleveref}
\usepackage{siunitx}
\usepackage[T1]{fontenc}
\usepackage{libertine}
\graphicspath{{images/}}
\usepackage{multirow}
\usepackage{gensymb}
\usepackage{verbatim}

\usepackage[utf8]{inputenc}
\usepackage[paper=a4paper,margin=1in]{geometry}
\usepackage[english]{babel}
\usepackage{xcolor}
\usepackage{graphicx}
\usepackage{float}
\usepackage{mwe}
\usepackage{xr}

\begin{document}

\title{Phase stability in the Hf-N and Zr-N systems}
\author{Jonathan Li}
\author{Derick Ober}
\author{Anton Van der Ven}
\email{avdv@ucsb.edu}
\affiliation{Materials Department, University of California Santa Barbara}
\date{\today}

\begin{abstract}
Hf and Zr nitrides are promising compounds for many technologically important areas, including high temperature structural applications, quantum computing and solar/optical applications.  
This article reports on a comprehensive first-principles statistical mechanics study of phase stability in the Hf-N and Zr-N binary systems. 
A high solubility of nitrogen in the hcp forms of Hf and Zr is predicted. 
The rocksalt forms of HfN and ZrN can also tolerate a high degree of off-stoichiometry through the introduction of nitrogen and metal vacancies. 
The Hf-N binary favors a family of stacking faulted parent crystal structures at intermediate nitrogen concentrations that host a unique form of short-range order among nitrogen interstitials and vacancies. 
These phases can accommodate some degree of configurational entropy and remain ordered to temperatures as high as 1200K.

\end{abstract}

\maketitle

\section{Introduction}
\label{sec:introduction}

Early transition metals, such as Ti, Zr and Hf, can dissolve high concentrations of carbon, nitrogen and oxygen, forming interstitial solid solutions and a variety of stoichiometric and substoichiometric compounds.\cite{pierson1996handbook,toth2014transition,puchala2013thermodynamics,gunda2018first,gunda2018resolving} 
The nitrides are a particularly interesting and increasingly important class of materials. 
Included in this group are the rocksalt mononitrides, such as ZrN and HfN, which have high melting temperatures and strengths that make them attractive materials for high temperature structural applications.\cite{fahrenholtz2014ultra,fahrenholtz2017ultra,wuchina2007uhtcs} 
Early transition metal nitrides are also used as nanoscale metal gate electrodes in transistors, making them compatible with silicon-based chip manufacturing technologies. \cite{bower2020complementary,seo2005epitaxial}
This combined with their low temperature superconducting properties \cite{cassinese2000transport} has made them of interest for quantum computing applications.\cite{potjan2023300} 
The optical properties of nitrides such as ZrN and HfN are also receiving attention, with potential applications as solar reflectors and as absorbers in hot carrier solar cells \cite{das2022refractory,chung2016hafnium} and as alternatives to silver and gold as plasmonic materials \cite{naik2013alternative}.

The binary Hf-N and Zr-N phase diagrams remain poorly characterized. 
Yet, many of the structural and functional properties of transition metal nitrides are sensitive to their nitrogen concentration, their crystal structure and the degree of ordering among their interstitial nitrogen atoms. 
The most recent experimental phase diagram for Hf-N, compiled by Okamoto \cite{okamoto1990hf}, shows an hcp phase ($\alpha$) with a N solubility up to 29 atom \% and a rocksalt form of HfN with a high melting temperature at nitrogen-rich compositions. 
Metallic Hf forms bcc ($\beta$) at higher temperatures, which is also able to dissolve an appreciable quantity of interstitial nitrogen. 
The Zr-N phase diagram \cite{okamoto2006n,ma2004thermodynamic} has much in common with that of Hf-N, also containing an hcp phase ($\alpha$) at dilute N compositions and a ZrN rocksalt with some tolerance for N vacancies. % at higher N compositions. 
Similar to Hf, a bcc ($\beta$) polymorph of Zr forms at high temperatures. 

A notable difference between the Hf-N and Zr-N phase diagrams is the stability of substoichiometric hafnium-nitides with compositions Hf$_3$N$_2$ and Hf$_4$N$_3$. 
These were discovered by Rudy et al \cite{rudy1970crystal}. 
Weinberger et al \cite{weinbergerzetaetahfn}, using first-principles density functional theory (DFT) calculations, predicted that the two phases consist of close-packed Hf layers with a periodic array of stacking faults, the $\eta$-Hf$_3$N$_2$ compound having an ABCBCACAB stacking (9R) of close-packed Hf atoms, and the $\zeta$-Hf$_4$N$_3$ compound having a ABCACABCBCAB stacking (12R) of close-packed Hf atoms. 
Both structures accommodate N atoms in their non-face sharing octahedral interstitial sites.
In the nitrogen-rich portion of the binary, Hf$_3$N$_4$ polymorphs have been observed experimentally \cite{zhang2014theoretical,kroll2003hafnium}.

The rocksalt phases in the Zr-N and Hf-N binaries, which have favorable high temperature structural properties and low temperature functional properties, can tolerate vacancies on both the nitrogen and the metal sublattices.\cite{yu2017first,balasubramanian2018energetics,zhang2017pressure,weinberger2017ab,ushakovnavrotskyavdw2019carbidesnitrides} 
The vacancies can order at low temperature to form a variety of nitride compounds with differing metal to nitrogen ratios.
Undoubtedly, optical properties and superconducting transition temperatures will be sensitive to the nitrogen composition and therefore to processing conditions.
At very high temperatures, the disorder among vacancies will contribute configurational entropy to the free energy and thereby affect high temperature phase stability. 
Very little is known about the ordering tendencies of vacancies in the Hf-N and Zr-N rocksalt phases and the degree to which vacancies are tolerated at high temperatures. 

In this article, we report on a comprehensive study of phase stability in the Hf-N and Zr-N binary systems using first-principles statistical mechanics methods.\cite{van2018first,puchala2023casm} 
We establish zero-Kelvin phase stability by systematically enumerating interstitial N orderings within likely parent crystal structures \cite{kolli2020discovering} and by calculating the energy of each structure with Density Functional Theory (DFT).  
Cluster expansion techniques \cite{sanchez1984generalized,de1994cluster,van2018first} are used to interpolate DFT energies for each parent crystal structure within Monte Carlo simulations, which are used to calculate finite temperature free energies that rigorously account for configurational disorder. 
The free energies are then used to calculate equilibrium temperature versus composition phase diagrams. 
Much of the intermediate analysis in this work will focus on the Hf-N system, as the Zr-N and Hf-N systems share many similarities.

\section{Methods}
\label{sec:methods}
%\subsubsection{Electronic structure calculations}
Density functional theory methods and statistical mechanics approaches were used to predict phase stability in the Hf-N and Zr-N binaries, both at 0K and at finite temperature. 
Density functional theory calculations were performed with the Vienna Ab-initio Simulation Package (VASP) \cite{kresse1993ab,kresse1994ab,kresse1996efficiency,kresse1996efficient} using the approximation of Perdew-Burke-Ernzerhof (PBE) \cite{perdew1996generalized} for the exchange-correlation functional.  
Interactions between core and valence electrons were treated with the Projector Augmented Wave (PAW) pseudopotential method \cite{blochl1994projector,kresse1999ultrasoft}. 
The PAW potentials treat the 4s, 4p, 5s, and 4d orbitals of Zr (Zr$_{sv}$); the 5p, 6s, and 5d orbitals of Hf (Hf$_{pv}$); and the 2s and 2p orbitals of N as valence states. 
The electronic structure was converged to a tolerance of 10$^{-5}$ eV, using a $\Gamma$-centered reciprocal space discretization of 45 k-points per \r{A}$^{-1}$ and a plane wave energy cutoff of 525 eV. 
Atomic geometries are converged to a maximum force tolerance of 0.02 eV/\r{A}. Static runs were performed on the relaxed structures using the tetrahedron method with Blöchl corrections \cite{blochl1994improved}. % to obtain more accurate energy values.\cite{blochl1994improved} 
All calculations were performed without spin polarization as we found a negligible difference in energy (<0.1meV/atom) between calculations of the ground state structures with and without spin polarization. 
Energy differences for the calculations of the ground states with and without spin polarization can be found in the Supporting Information. 
Bader charge \cite{bader1985atoms,bader1991quantum} calculations were performed using the scheme of Henkelman et. al. \cite{tang2009grid,sanville2007improved,henkelman2006fast,yu2011accurate}. 
The analysis of the octahedral distortion modes follows the algorithm described by Saber et al.\cite{saber2023redox}. 
%COHP calculations are performed using the LOBSTER software package \cite{deringer2011tchougreeff,deringer2011crystal,maintz2013analytic,maintz2016lobster,nelson2020lobster}. 

The enumeration of different nitrogen-vacancy and metal-vacancy orderings over various parent crystal structures was performed with the CASM software package \cite{CASM2023,puchala2023casm}. 
CASM was also used to construct and parameterize cluster expansion Hamiltonians and perform finite temperature Monte Carlo simulations. 
%\subsubsection{Fitting Cluster Expansion Hamiltonians}
DFT formation energies were used to fit cluster expansion Hamiltonians \cite{sanchez1984generalized,de1994cluster,van2018first} that parameterize the formation energies as a function of nitrogen(metal)-vacancy ordering. 
%The cluster expansions use occupational basis functions at each nitrogen(metal) site $i$ with value 1 if occupied and 0 if vacant.
Bayesian regression methods as described by Ober et al. \cite{ober2023thermodynamically} were used to parameterize cluster expansions that reproduce the same ground state orderings as predicted with DFT.
To ensure that anti-site disorder within each ordered ground state on the hcp and rocksalt parent structures is accurately reproduced by the cluster expansions, energies of point, pair and triplet anti-site defects within in large supercells of each ground state as enumerated with CASM were included in the DFT training data set used to fit the cluster expansions. 
%To improve the cluster expansion predictions of the ground states, CASM enumerated perturbation calculations were performed near each ground state in the $\alpha$ and $\delta$ phases.

%\subsubsection{Semi-grand Canonical Monte Carlo}
The cluster expansions for each parent crystal structure were used in semi-grand canonical Monte Carlo simulations to calculate free energies that rigorously account for configurational disorder. 
These free energies were then used to construct temperature-composition phase diagrams. 
The Monte Carlo simulations were conducted using the CASM software package over a range of temperatures and chemical potential with resolution of 20K and 5meV, respectively. \cite{puchala2023casmmonte}
Data processing and visualization was performed using CASM, \cite{puchala2023casm,CASM2023} the pymatgen package,\cite{ong2013python} VESTA software,\cite{momma2011vesta} and customized python scripts.
%Phase diagrams were then constructed using the common tangent method to determine two-phase regions. 

\section{Results}

\subsection{Electronic structure and interactions within dilute Hf-N and Zr-N}
\label{sec:electronic_structure}

\begin{figure}
    \centering
    \includegraphics[width=8cm]{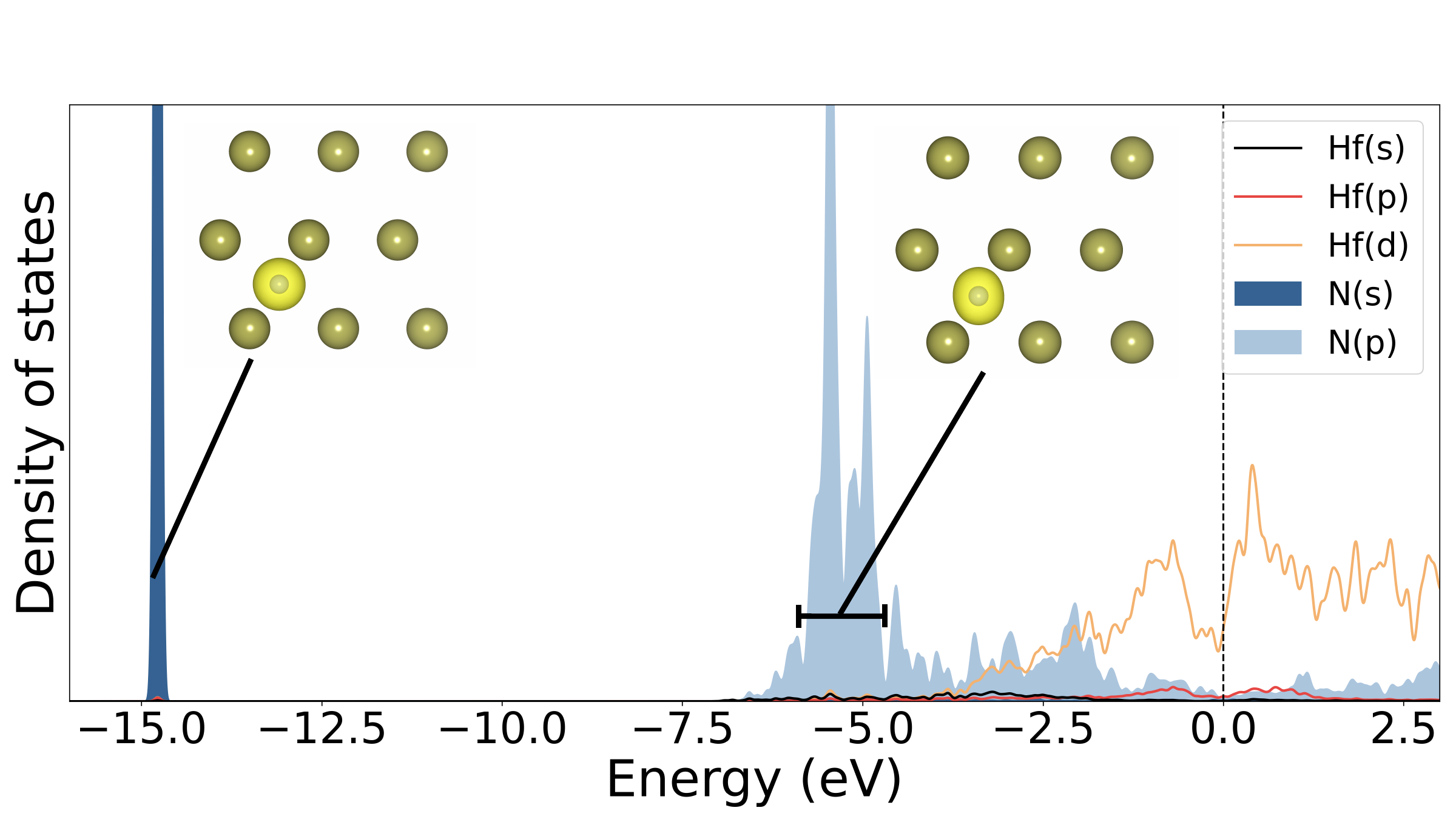}
    \caption{Density of states and partial charge densities for an octahedral N interstitial in a 5x5x3 supercell of the primitive hcp Hf unit cell. The partial charge densities highlight regions of interest in the N s- and p- orbitals.}
    \label{fig:1N_dos_parchg}
\end{figure}

Both Zr and Hf are able to dissolve very high concentrations of interstitial nitrogen.\cite{okamoto1990hf} % Hansen
Before analyzing phase stability in concentrated Zr-N and Hf-N compounds, however, we first consider interactions at dilute nitrogen concentrations. 
Figure \ref{fig:1N_dos_parchg} shows the electronic density of states of a large supercell of hcp Hf (a 5x5x3 supercell of the primitive hcp unit cell with 150 Hf atoms) containing an isolated interstitial nitrogen atom. 
As is evident in Figure \ref{fig:1N_dos_parchg}, both the nitrogen s and p levels are below the Hf d and s levels. 
The nitrogen s levels are highly localized as reflected by the sharp peak of its density of states. 
The nitrogen p levels are close to the bottom of the Hf d and s levels and exhibit some broadening due to a limited degree of hybridization with the Hf levels. 
However, in spite of this partial broadening, the nitrogen p levels remain localized. 
This can be seen upon inspection of the electronic charge distribution around the interstitial nitrogen corresponding to the states with energies in the interval in which the nitrogen p states have a finite DOS (Figure \ref{fig:1N_dos_parchg}). 
The p levels of a neutral nitrogen atom are only partially filled.
However, when nitrogen is dissolved in hcp Hf, the nitrogen p state energy levels are far below the Hf fermi level.
The Hf electrons fall into these lower energy states, meaning that a dissolved nitrogen atom is therefore locally charged.
This is confirmed by a calculation of the Bader charge, which for an isolated nitrogen atom in a 5x5x3 (150 Hf atoms) supercell of hcp Hf has a value of -1.9. 
Due to the high density of states at the Fermi level, the Hf host is metallic and the local charge accumulation around a nitrogen atom is screened by the itinerant electrons of the surrounding Hf host. 
Similar behavior is predicted for oxygen interstitials within Ti and Nb.\cite{gunda2021investigating,reynolds2024solute}

\begin{figure}[!htp]
    %\centering
    \includegraphics[width=\linewidth]{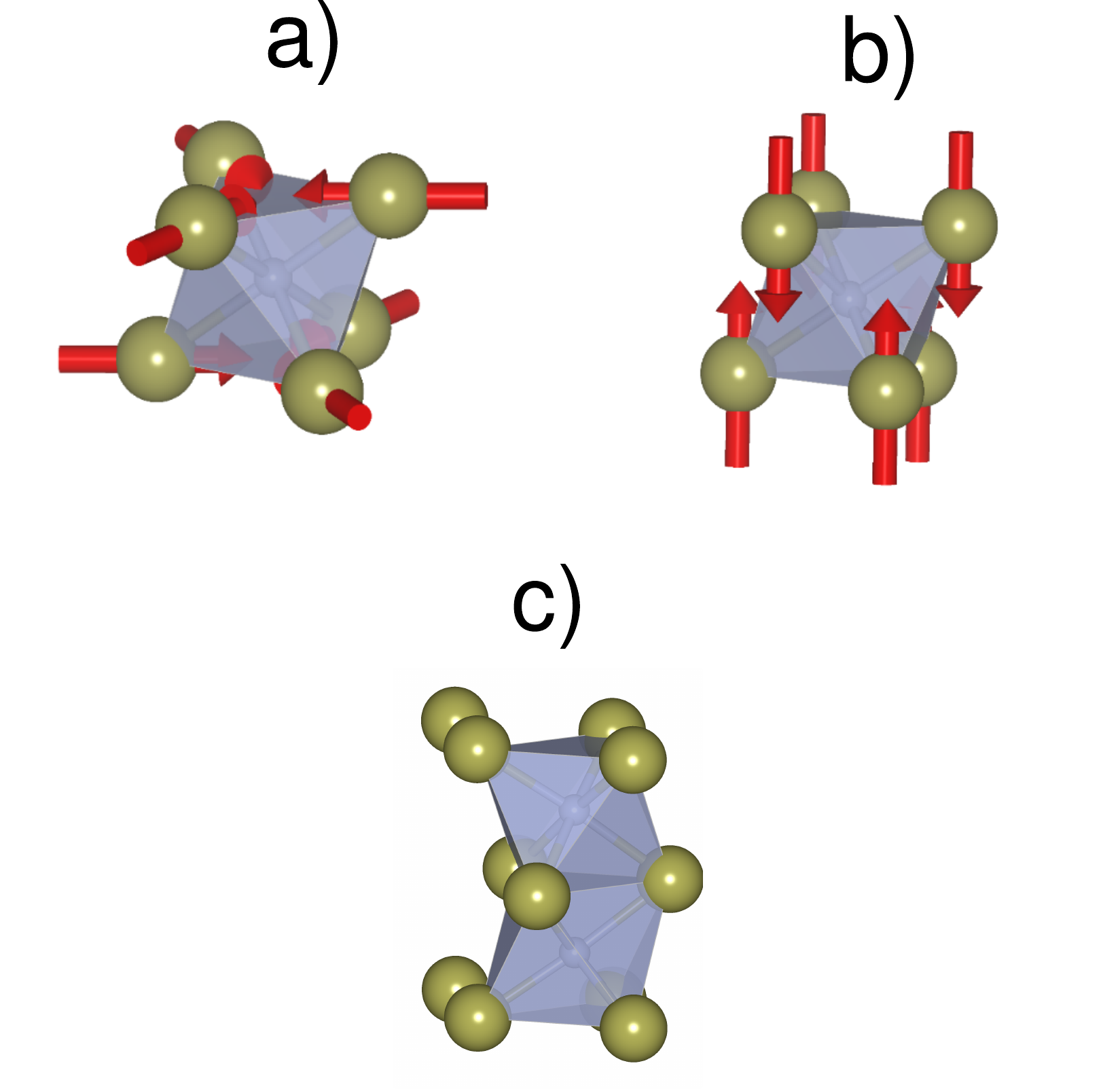}
    \caption{An interstitial nitrogen atom in hcp Hf or Zr will distort its coordinating metal atoms. Two symmetry adapted collective distortion modes of the octahedron of Hf/Zr atoms that coordinate a nitrogen interstitial have non-zero amplitudes. (a) The first collective distortion mode corresponds to a contraction or expansion of the octahedron within the close-packed planes of the hcp crystal. (b) The second collective distortion mode corresponds to a contraction or expansion of the distance between close-packed faces of the octahedron. (c) A representative pair of face-sharing octahedra in hcp Hf.}
    \label{fig:activated_irreps}
\end{figure}

Interstitial atoms will generally distort the surrounding crystal due to size misfit and changes in bonding. 
Local distortions can be tracked with symmetry adapted collective distortion modes of the cluster of metal atoms that coordinate the interstitial atom.\cite{thomas2013finite,saber2023redox}
To assess the degree with which an isolated nitrogen distorts the surrounding Hf or Zr crystal, we calculated the amplitudes of symmetry adapted distortion modes of the octahedron of Hf or Zr atoms coordinating an interstitial nitrogen atom. 
The collective distortion modes with non-zero amplitudes (relative to the undistorted octahedron in pristine hcp Hf or Zr) are shown in Figure \ref{fig:activated_irreps}.
The full set of distortion modes for an octahedron in hcp can be found in the Supporting Information.
The collective distortion modes are normalized to unity such that the amplitude of each mode measures the total distortion of the mode in units of \AA ngstrom.
%The distortion modes are calculated in reference to an equivalent octahedron in a perfect Hf/Zr hcp lattice.
%All non-zero normal mode amplitudes thus represent a deformation due to inserting an octahedral N interstitial. 
The amplitudes of the collective distortion modes of Figure \ref{fig:activated_irreps}(a) and (b) for an interstitial nitrogen in Hf (Zr) are 0.051 (0.063) \AA{} and -0.027 (-0.008) \AA{}, respectively.
The distortion in Figure \ref{fig:activated_irreps}(a) corresponds to a shrinking/expansion of the close packed triangular faces, while the distortion of Figure \ref{fig:activated_irreps}(b) corresponds to a contraction/expansion of the close-packed interlayer spacing, respectively.
In both Hf and Zr, an interstitial N atom results in a contraction of the close-packed triangular planes and a slight expansion along the z-axis.
The expansion along the z-axis is more pronounced in Hf than in Zr.

\begin{figure}[htbp]
  \includegraphics[width=\linewidth]{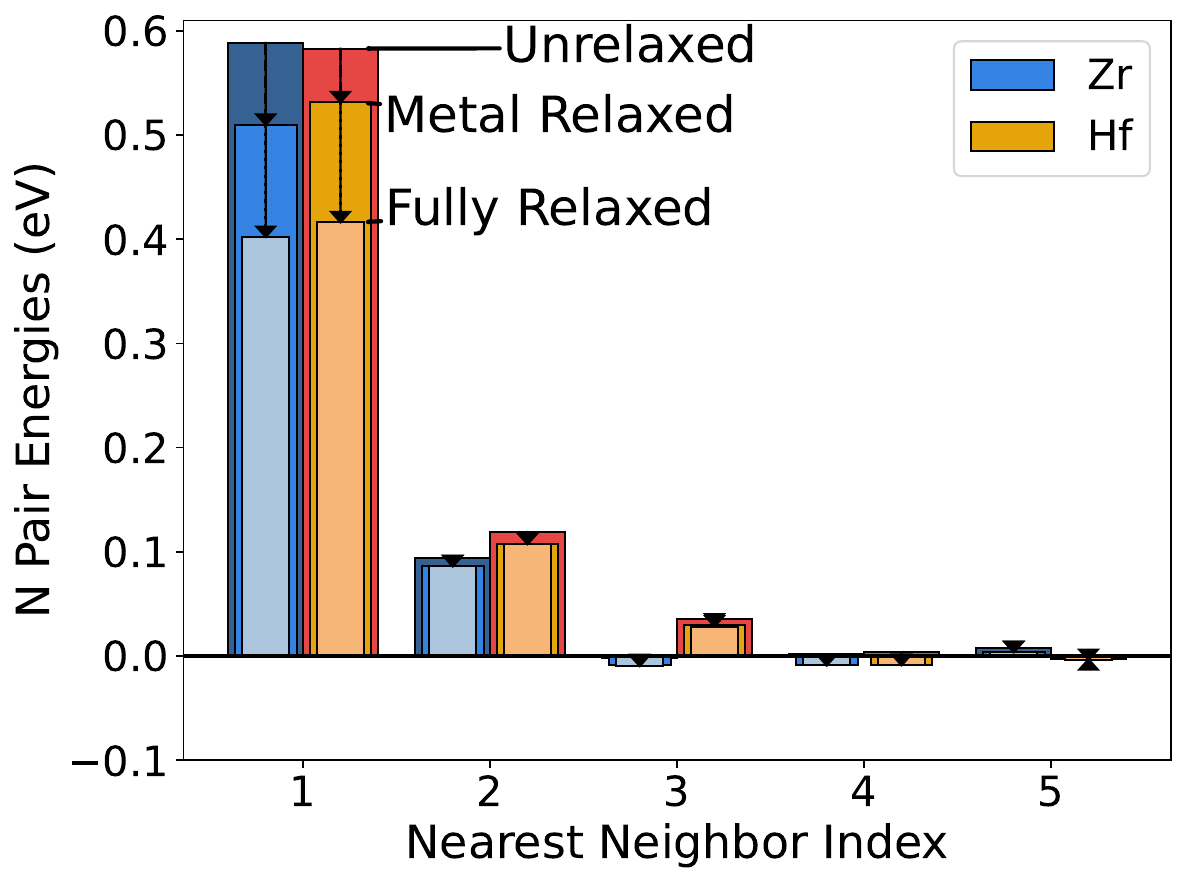}
  {\caption{Nitrogen-nitrogen pair energies in hcp Zr/Hf as a function of nearest neighbor pair index. Nearest neighbor index 1 corresponds to the first nearest neighbor pair, in which a pair of nitrogen atoms occupy face-sharing octahedra. Each nearest neighbor set shows the ideal, metal relaxed, and fully relaxed pair energies.}  
  \label{fig:N-N_pair_energies}}
\end{figure}

We next consider a pair of nitrogen atoms in a 5x5x3 supercell of hcp Hf and Zr. 
Figure \ref{fig:N-N_pair_energies} shows pair energies for interstitial N atoms in hcp Hf and Zr as a function of distance, and for different degrees of relaxation.
We first considered pair energies in the absence of any relaxations. 
The interstitial N atoms were placed at the center of their octahedra and the metal atoms maintained the same positions as in the pure hcp crystal. 
These pair energies in Figure \ref{fig:N-N_pair_energies} are shown as the bars with the darkest color. 
We next considered pair energies where only the surrounding Hf or Zr were allowed to relax. 
In this second set of calculations, the nitrogen atoms were held fixed at the center of their octahedra. 
In the third set of calculations, all atoms were allowed to fully relax (metals and nitrogen).
These three scenarios allow us to separate out the role of electronic and electrostatic interactions from those mediated by elastic deformations of the crystal.
%Pair energies are calculated for the unrelaxed state - where the interstitial N atoms are placed at ideal octahedral sites on the hcp metal lattice, metal relaxed state - where only the metal lattice is allowed to relax, thus maintaining a fixed N-N distance, and the fully relaxed state.
The energies of each supercell calculation are referenced to the energy of two isolated nitrogen interstitials, where positive pair energies indicate repulsive interactions.
The fourth and fifth nearest neighbor pair energies are small (<5meV) and indicate that there is minimal interaction between the N interstitials beyond the third nearest neighbor shell. 

As is clear in Figure \ref{fig:N-N_pair_energies}, the first nearest neighbor interaction is highly unfavorable. 
This configuration corresponds to a pair of nitrogen atoms occupying face-sharing octahedral sites of adjacent layers as depicted in Figure \ref{fig:activated_irreps}(c).
The pair energy drops dramatically when the pair of nitrogen atoms go from a first nearest neighbor configuration to a second nearest neighbor configuration, with less of a decrease as the pair distance increases further. 
Figure \ref{fig:N-N_pair_energies} shows a significant decrease in the nearest neighbor pair interaction when the metal atoms are allowed to relax. 
Since the pair interaction energy is relative to a pair of isolated nitrogen atoms, this reduction indicates that the energy gain due to metal relaxations around each nitrogen interstitial is larger when the nitrogen atoms are nearest neighbor pairs than when they are separated by a large distance.   
The change in pair energy from the unrelaxed state to the metal relaxed state is only significant for the first nearest neighbor configuration, indicating that the effect of a pair of interstitials on the metal lattice is similar for any pair from the second nearest neighbor outwards. 
%In the first nearest neighbor case, the pair energy from unrelaxed to the metal relaxed state relaxes more than the for the fifth nearest neighbor case.
%This indicates that the relaxations of the lattice in the first nearest neighbor case is favorable relative to two isolated N interstitials.
When the nitrogen atoms are also allowed to relax, there is a further gain in energy that is most pronounced for the nearest neighbor pair configuration. 
This is because the metal octahedra surrounding the nitrogen interstitial atoms of a nearest neighbor pair share a face. 
The nitrogen atoms lower their energy by relaxing away from each other. 
%A similar trend can be seen from the metal relaxed state to the fully relaxed state.
%This can be simply understood by the first nearest neighbor N-N pairs moving apart due to electrostatic repulsion to lower their energy, which does not occur for further nearest neighbor pairs.

\begin{figure}
    \centering
    \includegraphics[width=8cm]{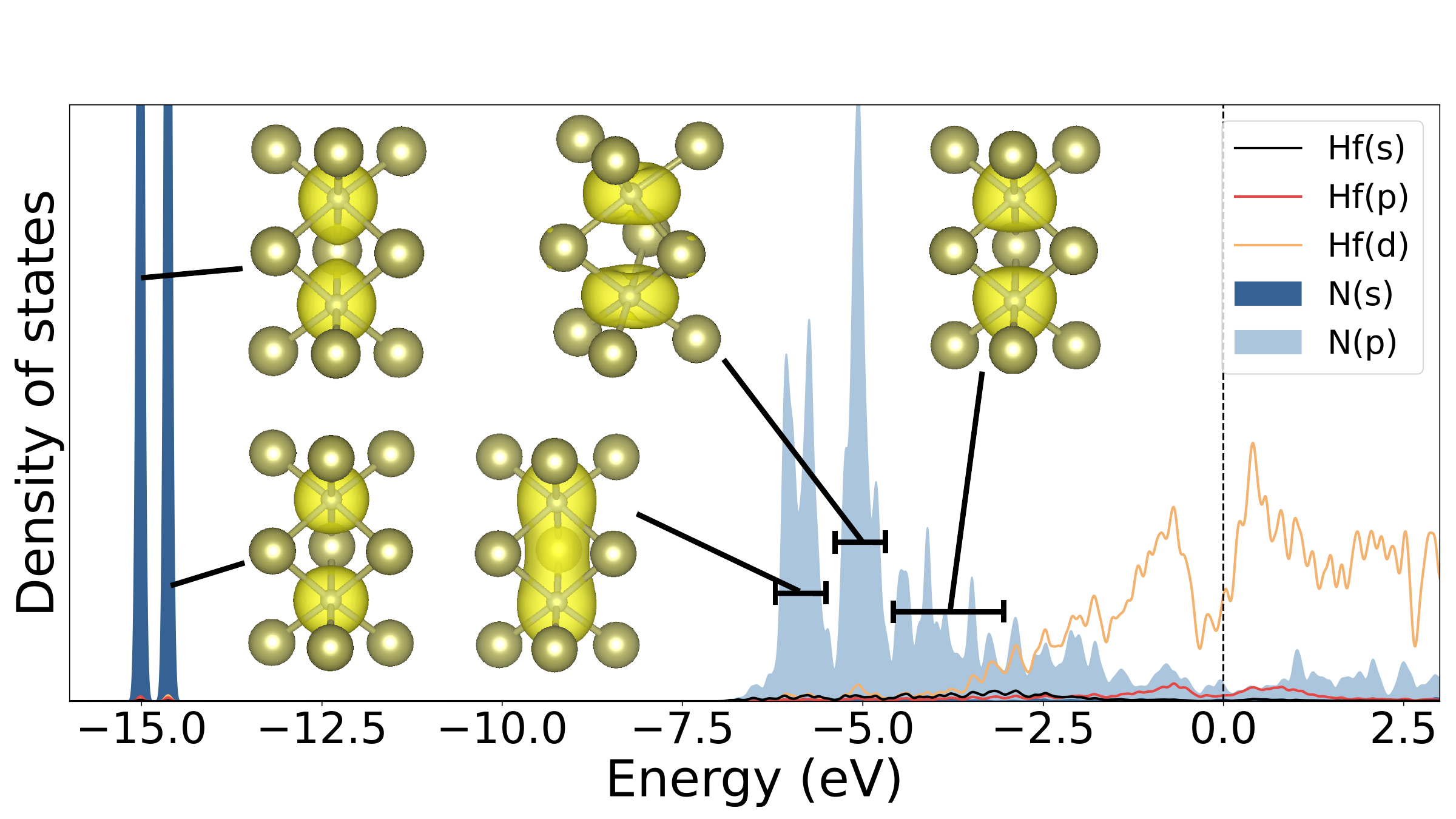}
    \caption{Density of states and partial charge densities for 1st nearest neighbor N interstitials in a 5x5x3 supercell of the primitive hcp Hf unit cell. The partial charge densities show the N s- and p-orbital splitting into bonding, non-bonding, and anti-bonding states. }
    \label{fig:1st_NN_dos_parchg}
\end{figure}

There are several ways in which neighboring interstitial species within early transition metals interact with each other.\cite{gunda2020understanding,gunda2021investigating,reynolds2024solute}
Figure \ref{fig:1st_NN_dos_parchg} shows the electronic density of states of a pair of nearest neighbor nitrogen atoms in hcp Hf. 
The s orbitals of the nearest neighbor nitrogen atoms overlap to some extent and hybridize to form bonding and antibonding orbitals as is evident in the DOS plot and the accompanying charge density plots in Figure \ref{fig:1st_NN_dos_parchg}. 
Similarly, a pair of p-orbitals aligned parallel to the c-axis of the  hcp unit cell form a $\sigma$ bonding and a $\sigma$ anti-bonding orbital, while the p-orbitals with lobes parallel to the basal plane of the hcp unit cell form bonding and antibonding $\pi$ orbitals. 
Since all nitrogen s and p orbitals are far below the Fermi level of the Hf host, they are completely filled and any hybridization between these levels results in completely filled bonding and anti-bonding states. 
There are, therefore, no favorable chemical interactions arising from the hybridization between localized nitrogen orbitals that would lead to an attraction between nearest neighbor nitrogen atoms.

\begin{figure}[htbp]
    \centering
    \includegraphics[width=7.5cm]{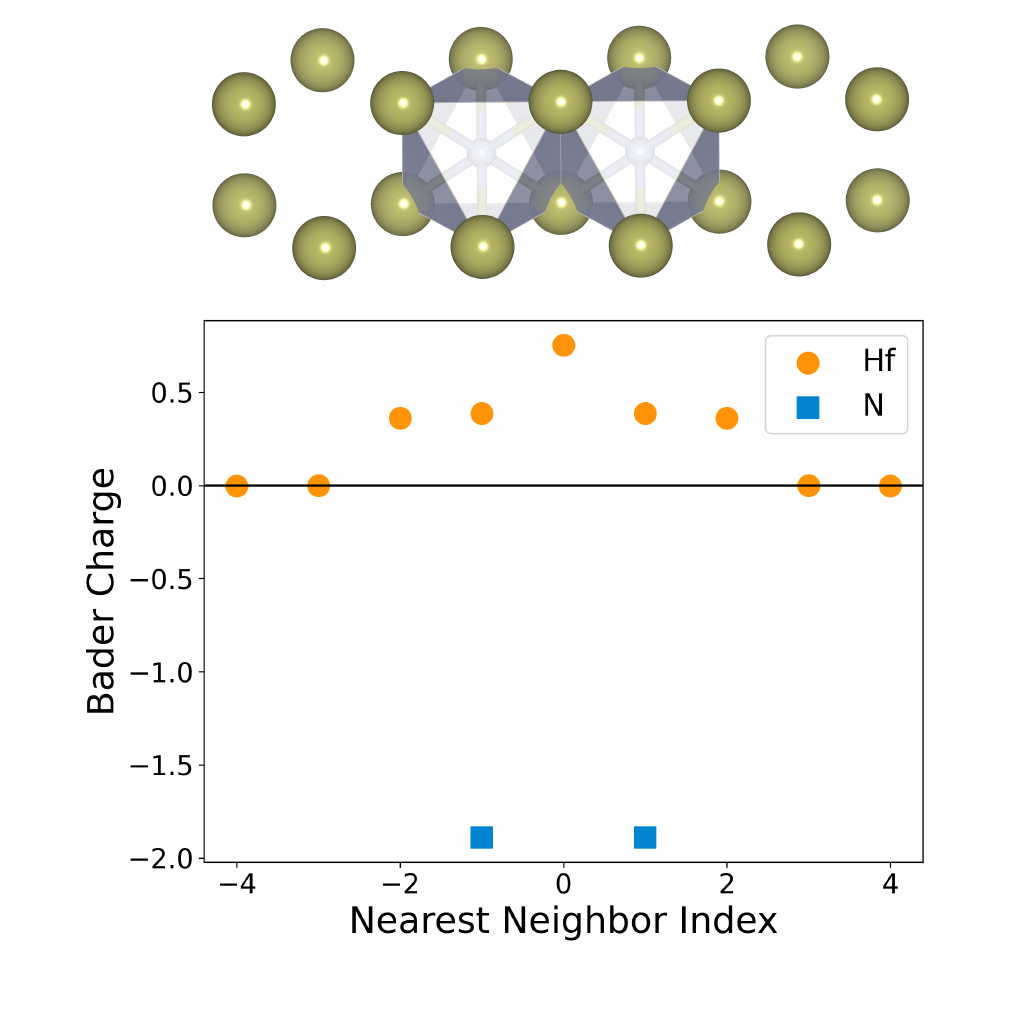}
    \caption{Bader charge in hcp Hf with 2nd nearest neighbor N interstitials. Positive Bader charge represents net loss of electron charge. The in-place view for the 2nd nearest neighbor N interstitials is shown.}
    \label{fig:bader_charge_2nd_NN_N-N}
\end{figure}

The pair of interstitial nitrogen atoms can still interact by other means.
Due to their local negative charge, they can exert an electrostatic repulsion on each other at short distances.\cite{gunda2020understanding,gunda2018first,reynolds2024solute}
%The repulsive interaction derives primarily from electrostatic forces between a pair of negatively charged nitrogen. 
Since a nearest neighbor pair of nitrogen atoms within hcp occupy face-sharing octahedra, there are no intervening Hf atoms to screen electrostatic interactions. 
When a pair of nitrogen atoms occupy non-face-sharing octahedra, as in a second-nearest neighbor configuration, the local negative charges on the nitrogen atoms are more effectively screened by Hf atoms. 
This is evident in Figure \ref{fig:bader_charge_2nd_NN_N-N}, which shows the Bader charge for a pair of nitrogen interstitials at a second nearest neighbor spacing in a large (36 atom) hcp host lattice. 
Since hcp Hf with dilute interstitial N is metallic, the electron charge transfer from the metallic host to the nitrogen atoms is localized to Hf atoms within the 1st nearest neighbor shell of the interstitial N atom. 
Each Hf atom in the surrounding shell has positive Bader charge proportional to the number of coordinated N interstitials. 
%The strong repulsion of the first nearest neighbor N pairs in hcp follow expectations from Pauling's rules since they would result in unfavorable configurations with face-sharing octahedra and appears to be a driving force in the stability of these ground state configurations in the Hf-N system.

\subsection{Zero Kelvin phase stability}
\label{sec:zero_kelvin_phase_stability}

\begin{figure}[htbp]
    \includegraphics[width=7cm]{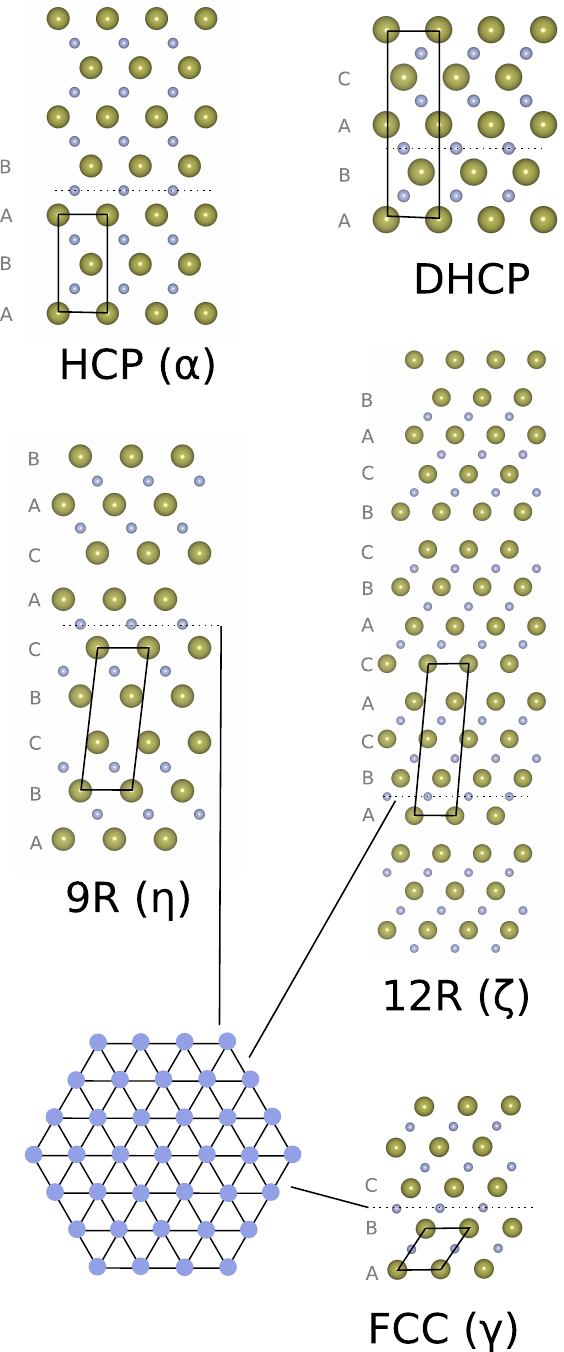}
    \caption{Stacking arrangements of the close-packed layers  in hcp, dhcp, 9R ($\eta$), 12R ($\zeta$), and rocksalt phases. Primitive unit cells are highlighted for each crystal system. 9R and 12R phases are shown with empty N layers at the stacking fault.}
    \label{fig:hcp-sigma-rs_image}
\end{figure}

The short-range repulsion between nearest-neighbor nitrogen atoms occupying face-sharing octahedra has important consequences for phase stability at non-dilute nitrogen concentrations. 
Each phase that has been observed or predicted to be stable in the Hf-N and Zr-N binary systems can be described as a particular metal crystal structure that hosts nitrogen atoms within their octahedral interstitial sites. \cite{ushakovnavrotskyavdw2019carbidesnitrides} 
This includes the hcp and fcc crystal structures as well as different close-packed structures such as double hcp (dhcp), having an ABAC stacking sequence of close-packed metal planes, 9R having an ABCBCACAB stacking and 12R having an ABCACABCBCAB stacking (Figure \ref{fig:hcp-sigma-rs_image}).
%Each metal host structure can accommodate nitrogen atoms within octahedrally coordinated interstitial sites. 
%In hcp, octahedral sites of adjacent layers share faces. 
As shown in the previous section, nearest neighbor pairs of interstitial atoms in hcp, which simultaneously occupy face-sharing octahedral sites, exert a strong repulsive force on each other. 
Configurations in which pairs of nitrogen fill face-sharing octahedra are therefore very unfavorable. 

In an fcc metal host structure, in the octahedral interstitial sites only share edges and interstitial nitrogen atoms in nearest neighbor octahedral sites are separated by a larger distance than in hcp. 
The fcc host can therefore accommodate more interstitial nitrogen per Hf than the hcp host without incurring the short-range repulsive interactions between face-sharing octahedra.
The stacking faulted hybrid host structures, such as 9R and 12R, combine slabs of fcc of varying thickness interleaved with stacking faults that have a local hcp environment.
The octahedral sites within the fcc slabs are non-face sharing, but the octahedral sites within the stacking fault share faces with octahedral sites below and above the stacking fault. 
In the hybrid hosts, therefore, not all octahedral sites can be filled with nitrogen without filling face-sharing octahedral sites. 

To establish phase stability at zero Kelvin in the Hf-N and Zr-N binary systems, we calculated the fully relaxed DFT-PBE energies of large numbers of ordered arrangements of nitrogen and vacancies over the octahedral interstitial sites of the hcp, fcc, dhcp, 6R, 9R, 12R, 15R (5R), 18R and 21R metal host structures. 
The coordinates of each parent crystal structure are collected in the Supporting Information. 
Consistent with past literature, we will refer to the hcp form of Zr and Hf as the $\alpha$ phase and the fcc-derived rocksalt as the $\delta$ phase. 
Following Weinberger et al \cite{weinbergerzetaetahfn}, phases generated upon the insertion of nitrogen in the octahedral interstitial sites of 9R (12R) will be referred to as $\eta$ ($\zeta$). 
Table \ref{tab:calculated_structures} lists the number of nitrogen-vacancy orderings over the interstitial sites of different parent crystal structures whose energies were calculated with DFT-PBE. 

\begin{table*}[]
    \centering
    \begin{tabular}{c c c c} \hline \hline
         Crystal Structure & Close-packed Motif & Orderings Calculated (Hf) & Orderings Calculated (Zr) \\ \hline
         hcp & AB & 723 & 1560 \\ 
         6R & ABACBC & 13 & 14 \\ 
         9R ($\eta$) & ABCBCACAB & 356 & 16 \\ 
         12R ($\zeta$) & ABCACABCBCAB & 10 & 8 \\ 
         15R (5R) & ABCBC & 13 & 13 \\ 
         18R & ABCABCBCABCACABCAB & 1 & N/A \\ 
         21R & ABCABCACABCABCBCABCAB & 1 & N/A \\ 
         dhcp & ABAC & 102 & 5 \\ 
         fcc & ABC & 1952 & 1699 \\ 
         %704+380+868 (Hf)
         %1297+402 (Zr)
         Misc. & N/A & 39 & 35 \\ \hline \hline
    \end{tabular}
    \caption{Summary of the number of nitrogen-vacancy orderings over the octahedral interstitial sites of different parent crystal structures whose energies were calculated with DFT-PBE. The energies of all low energy structures in the Zr-N binary system were also calculated for the Hf-N binary and vice versa.}
    \label{tab:calculated_structures}
\end{table*}

The energies of all symmetrically distinct nitrogen-vacancy configurations in all symmetrically distinct supercells containing up to 8 metal sites in both hcp and fcc phases were calculated. 
The energies of configurations in supercells containing more than 8 metal sites were calculated with DFT-PBE when their predicted energies using a cluster expansion were below or close to a tentative convex hull. 
Cluster expansions were iteratively refined until no new ground states were predicted. 
To the best of our knowledge, this is the most complete computational survey of the hcp and fcc hosts in the Hf-N binary to date, with calculations of the energies of structures containing as many as 32 metal sites.

\begin{figure}[htbp]
    \centering
        \includegraphics[width=\linewidth]{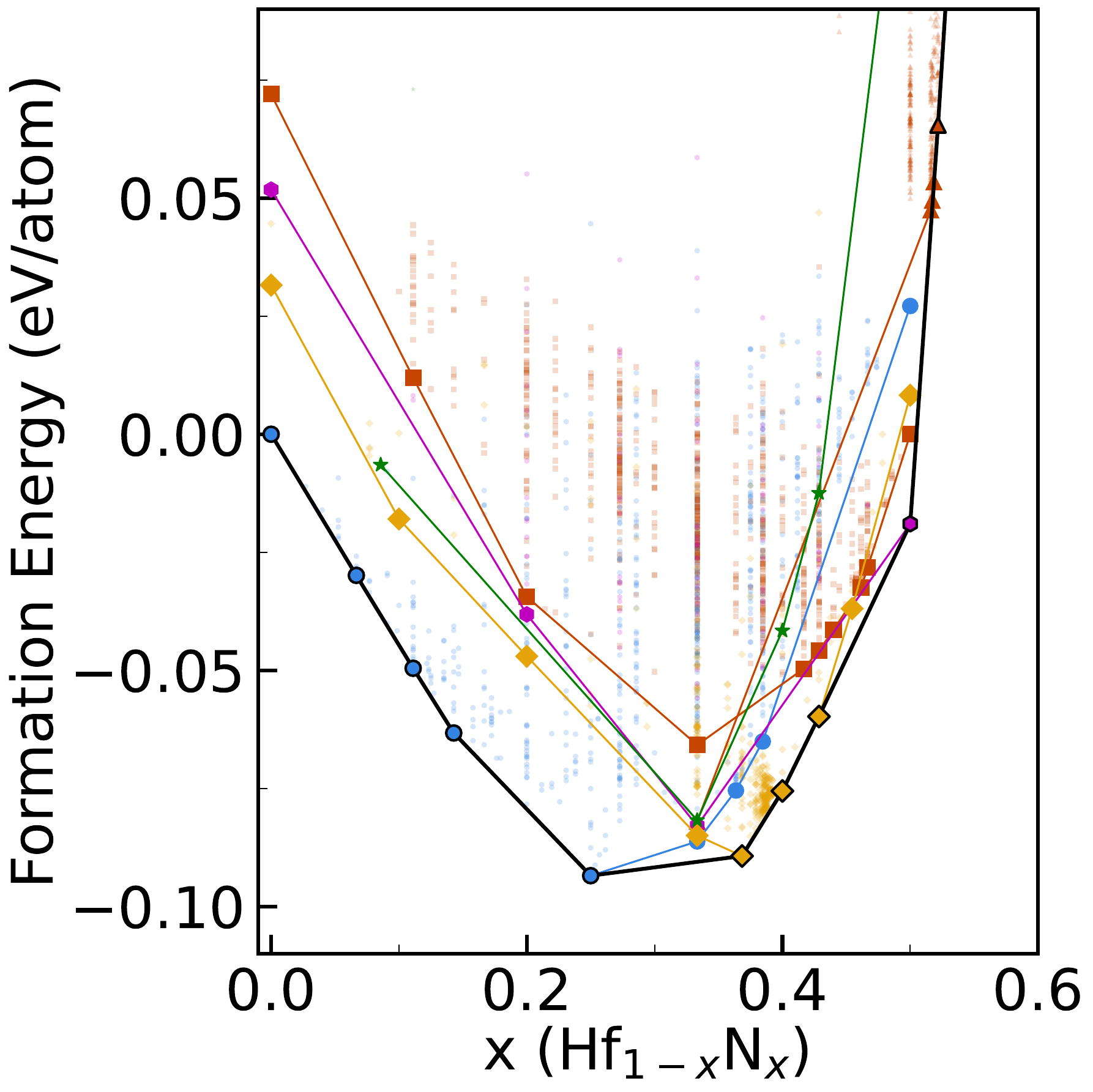}
        \caption{Formation energies in the Hf-N binary system as calculated with DFT-PBE. The global convex hull is shown in black. Individual convex hulls for each parent crystal structure are shown in their corresponding colors.}
        \label{fig:hf-n_hull}
        \includegraphics[width=\linewidth]{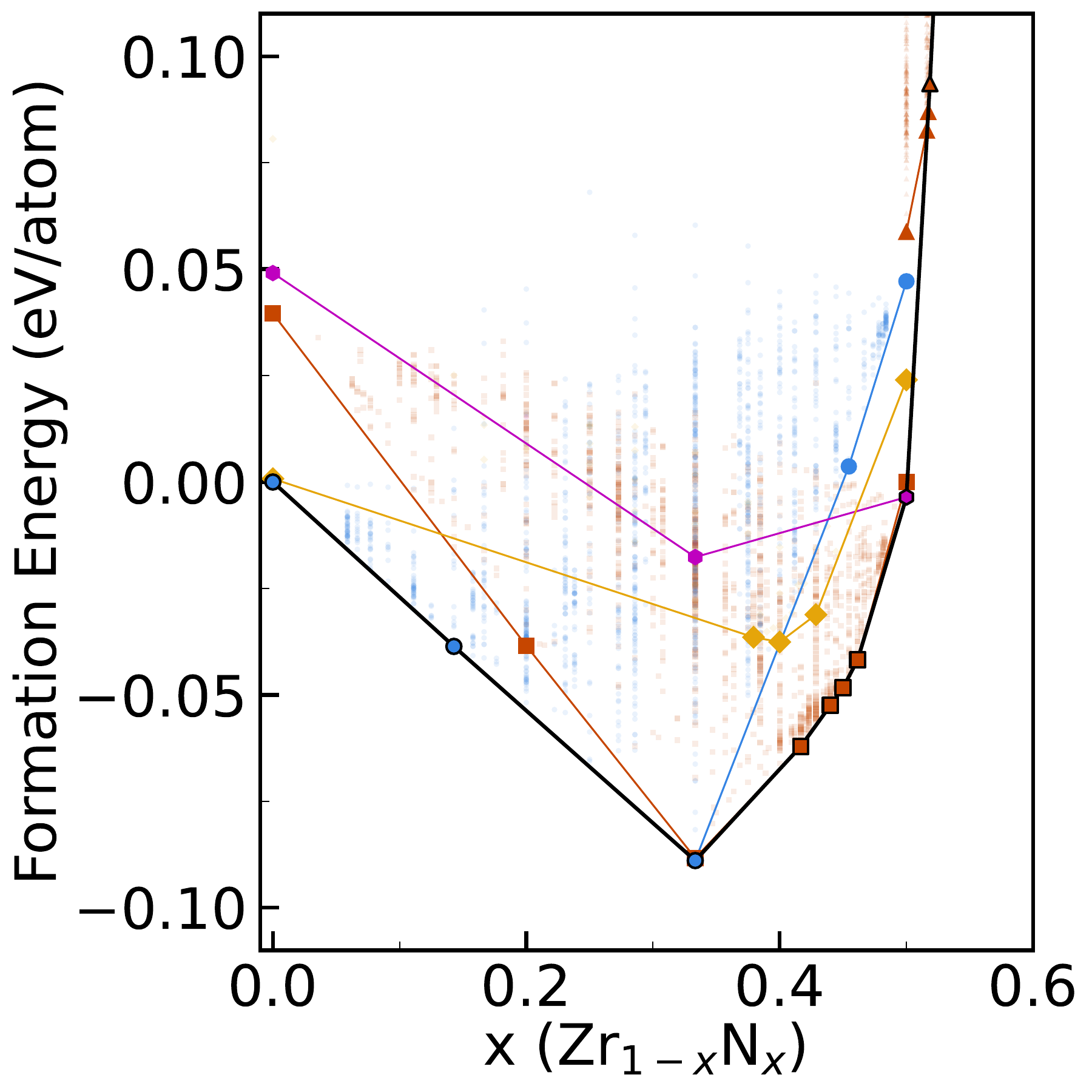}
        \includegraphics[width=\linewidth]{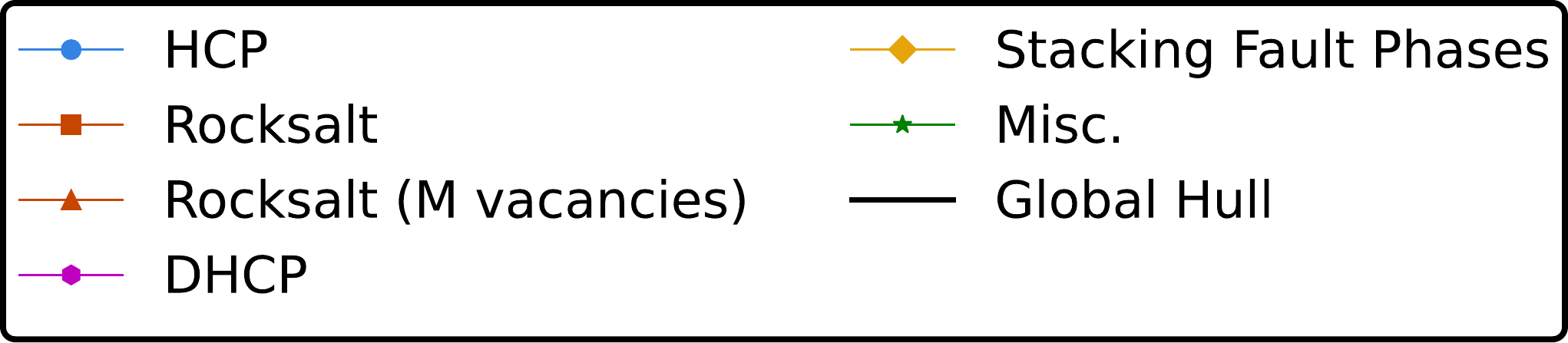}
        \caption{Formation energies in the Zr-N binary system as calculated with DFT-PBE. The global convex hull is shown in black. Individual convex hulls for each parent crystal structure are shown in their corresponding colors.}
        \label{fig:zr-n hull}
\end{figure}

Figures \ref{fig:hf-n_hull} and \ref{fig:zr-n hull} show the calculated formation energies for the Hf-N and Zr-N binaries, respectively. 
The formation energies in Figures \ref{fig:hf-n_hull} (Figure \ref{fig:zr-n hull}) were calculated using the energy of hcp Hf (Zr) and the chemical potential of nitrogen gas, $\mu_{N_2}$, at room temperature as reference states according to
\begin{equation}
\label{eqn:formation_energy}
    \Delta e_{M_{1-x}N_x} = e_{M_{1-x}N_x}^{DFT} - (1-x) e_{M}^{hcp} - x\mu_{N}     
\end{equation}
In this expression, $x$ is the nitrogen atom fraction, $M$ refers to Hf or Zr and the DFT energy of each configuration, $e_{M_{1-x}N_x}^{DFT}$, is normalized by the number of atoms in the unit cell. 
The nitrogen chemical potential, $\mu_{N}$, is equal to $1/2\mu_{N_2}(T=298K)$, the chemical potential of N$_2$ gas at room temperature.
We performed a sensitivity analysis of the nitrogen solubility in rocksalt when in equilibrium with a N$_2$ gas due to perturbations to $\mu_N$ and determined that the phase boundaries are fairly insensitive when $\mu_N$ is shifted by 0.5eV/N, approximately the value of PBE corrections for the O$_2$ binding energy as described by Wang et al. \cite{wang2006oxidation}.
Thus, we did not include an additional correction to $\mu_N$ to account for the PBE errors in calculating the binding energy of the N$_2$ molecule.
The analytical expression for the temperature dependence of $\mu_N$, and the sensitivity analysis with respect to the N$_2$ binding energy correction can be found in the Supporting Information.

\begin{figure}[htbp]
    \centering
    \includegraphics[width=\linewidth]{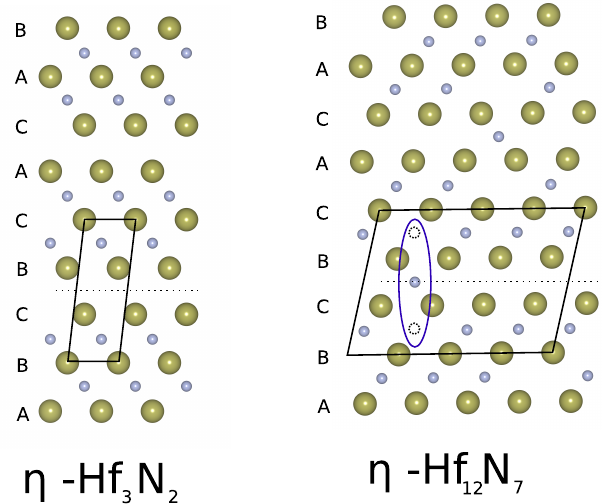}
    \caption{The crystal structures of $\eta$-Hf$_3$N$_2$ (left) and $\eta$-Hf$_{12}$N$_7$ (right). The dumbbell motif in $\eta$-Hf$_{12}$N$_7$ is circled in blue. Dashed circle represents nitrogen vacancies and dashed line represents the nitrogen vacant stacking fault plane.}  
    \label{fig:hf12n7_ordering}
\end{figure}

The convex hull to the formation energies of the Hf-N binary in Figure \ref{fig:hf-n_hull} shows that hcp Hf is stable up to a nitrogen atom fraction of $x$=0.25. 
Several ordered arrangements of nitrogen over the octahedral interstitial sites of hcp Hf are predicted to reside on the convex hull. 
A subset of the stacking faulted hybrid phases are predicted to be stable at higher nitrogen compositions. 
For each hybrid phase, configurations with unfilled N layers along the stacking fault have the lowest energies. 
These configurations have all non-face-sharing octahedral sites within the fcc slabs occupied and avoid the simultaneous occupancy of face-sharing octahedra at the stacking faults. 
Of these low energy configurations, only $\eta$-Hf$_{3}$N$_2$ in the 9R host and $\zeta$-Hf$_{4}$N$_3$ in the 12R host are on the global convex hull. 
This result is consistent with the calculations of Weinberger et al.\cite{weinbergerzetaetahfn}
Figure \ref{fig:hf12n7_ordering}(a) shows the filling of interstitial sites in the Hf$_{3}$N$_2$ ground state. 
Another 9R configuration, having stoichiometry Hf$_{12}$N$_7$ is also predicted to be on the global convex hull (Figure \ref{fig:hf12n7_ordering}(b)). 
This structure, discovered by Weinberger \cite{weinbergerzetaetahfn} and labeled $\eta$-Hf$_{12}$N$_7$, has a unique nitrogen ordering relative to the other stable orderings in the hybrid phases.
It contains a dilute nitrogen interstitial in the otherwise vacant stacking fault plane, but has vacancies in the face-sharing octahedral sites of the adjacent fcc blocks above and below the stacking fault.   
This configuration of a nitrogen interstitial and two adjacent nitrogen vacancies forms a "dumbbell" motif that can be tiled in 9R to form more candidate ground state structures. 
We systematically enumerated arrangements of this dumbbell motif in 9R supercells containing up to 48 Hf atoms (16 primitive cells of $\eta$-Hf$_3$N$_2$) and calculated their energies. 
None of these dumbbell motif arrangements were new ground states relative to $\eta$-Hf$_{12}$N$_7$ and $\eta$-Hf$_3$N$_2$.
Nevertheless, some arrangements have energies near the convex hull (<2meV/atom), indicating that configurational entropy may stabilize disordered arrangements of dumbbells at finite temperature.
% so we did not extend the search further due to computational costs.

While the $\delta$ phase (rocksalt) is experimentally observed at the HfN composition, our calculations predict dhcp as being the ground state of stoichiometric HfN, which aligns with previous work by Zhang, et al. \cite{zhang2017pressure}. 
Calculations of the vibrational free energies of rocksalt and dhcp HfN predict that the rocksalt phase becomes stable above 670 K.\cite{zhang2017pressure}
Hence the convex hull applied to the formation energies of the fcc-derived configurations (i.e. rocksalt) in the absence of the dhcp form of HfN sheds light on what can be expected at temperatures above 670 K.

Previous experimental evidence suggests that HfN rocksalt can accommodate up to 12.63\% vacancies on both the Hf and N sublattices \cite{straumanis1967unvollkommene}. 
We calculated energies near the stoichiometric composition with up to 1/3 vacancies on both the Hf and N sites. 
Rocksalt structures with vacancies on both sublattices are less stable than configurations with vacancies on only the N(Hf) sublattice at sub(super)nitride compositions. 
At the stoichiometric composition, all structures with vacancies on both sublattices have energies that are greater than 50meV/atom above that of the vacancy-free rocksalt structure. 
%The work by Straumanis that observed 12.63\% vacancies on both Hf and N sites may have instead observed a supernitride ground state with 13.62\% Hf vacancies and no N vacancies, which is plausible as there are multiple ground state, and near ground state Hf-vacancy ordered rocksalts at that composition (x$\approx$1.16). 
In the nitrogen rich (x>1) regime, the only stable ordered phases are rocksalt orderings with vacancies on the Hf-sublattice. 
We calculated supernitride rocksalt structures with Hf-vacancies up to 16 Hf sites. 
These supernitride phases have very shallow stability windows in the dilute Hf vacancy regime and there are many configurations that have energies within 1meV/atom of the convex hull, which is within the range of DFT error. 
The complete list of 0K ground state configurations is found in the Supporting Information . 
%The high number of Hf-vacancy ordered rocksalt (near) ground states suggests that there may be additional 0K ground states with Hf-vacancies in higher volume cells, but the exponential increase in possible arrangements with number of sites makes a brute force search unreasonable. 
%Additionally, the very shallow stability windows of these ground states suggests that the these ordered phases will disorder at low temperature, which is discussed below.

The convex hull to the formation energies of the Zr-N binary in Figure \ref{fig:zr-n hull} is qualitatively different from that of Hf-N (Figure \ref{fig:hf-n_hull}). 
While hcp Zr is stable at low nitrogen compositions and rocksalt Zr$_{1-x}$N$_x$ is stable at higher nitrogen compositions, none of the hybrid phases such as $\eta$ and $\zeta$ are predicted to be stable. 
At $x$=0.33 (corresponding to Hf$_2$N), there are nitrogen-vacancy orderings within hcp and fcc that are degenerate in energy.

\begin{figure}[H]
    \centering
    \includegraphics[width=8cm]{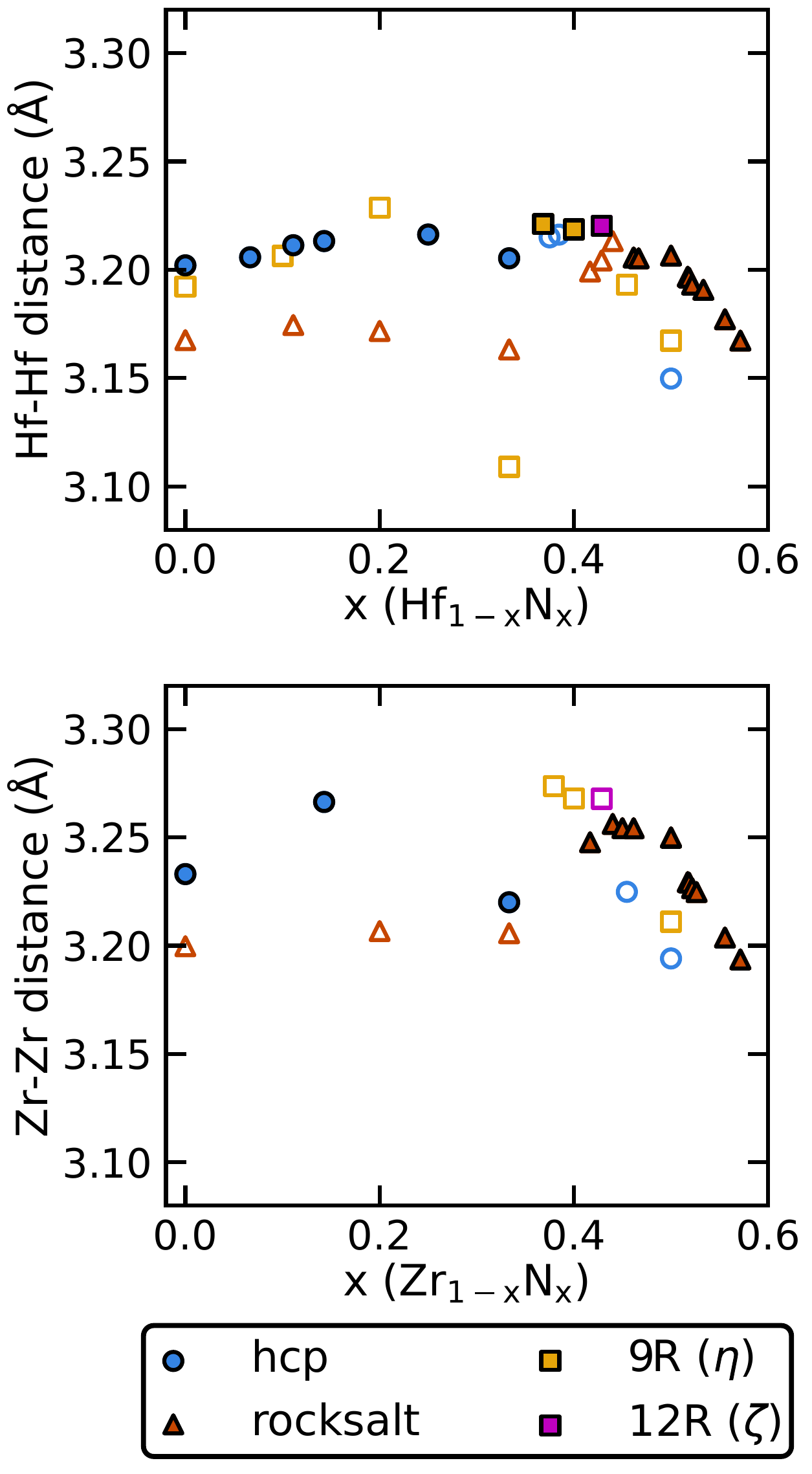}
    \caption{In-plane Hf-Hf (Zr-Zr) distance in the local ground states of Hf-N (top) and Zr-N (bottom) binary systems. Filled data points represent ground states also on the global convex hull.}
    \label{fig:in-plane_lattice_params}
\end{figure}

The absence of the $\eta$ and $\zeta$ stacking fault phases in the Zr-N binary can be attributed to epitaxial lattice strains. 
Figure \ref{fig:in-plane_lattice_params} shows the metal-metal distance within the close packed planes as a function of composition for the local ground states in each crystal system. 
In the Hf-N system, the Hf-Hf distance changes negligibly from hcp to 9R to 12R to rocksalt in the composition ranges where each phase is stable. 
Notably, the difference in Hf-Hf distances in $\alpha$-Hf, $\alpha$-Hf$_2$N, $\eta$-Hf$_3$N$_2$, $\zeta$-Hf$_4$N$_3$, and $\delta$-HfN is less than 1\%, indicating that the epitaxial strain energy that emerges when combining hcp and rocksalt slabs to form the 9R and 12R phases is minimal. 
In contrast, the Zr-Zr distance changes significantly with composition when moving from hcp to 9R to 12R to rocksalt. 
The difference in Zr-Zr distance between $\alpha$-Zr, $\alpha$-Zr$_2$N, $\eta$-Zr$_3$N$_2$, $\zeta$-Zr$_4$N$_3$, and $\delta$-ZrN is greater than 2\%, resulting in a significant epitaxial strain energy cost when hcp and rocksalt slabs are combined to form the 9R and 12R phases. % orderings as the hcp and rocksalt type layers would need to strain significantly to form the $\eta$ and $\zeta$ orderings. 
%Thus, Zr-N does not facilitate stability of the stacking fault phases.

\subsection{The DHCP Anomaly}
\label{sec:DCHP_anomaly}

The crystal structures of the ground states of the Hf-N binary show a clear trend as a function of the nitrogen concentration, with an increase in the N concentration inducing a change of the stacking sequence of the close-packed Hf layers. 
In large part, this occurs to avoid the simultaneous occupancy of face-sharing octahedral sites.
As shown in Section \ref{sec:electronic_structure}, pairs of N interstitials in dilute hcp Hf avoid face-sharing octahedral sites due to unfavorable electrostatic interactions between N atoms at short distances. 
Table \ref{tab:xtal_struct_densities} collects the fraction of face-sharing octahedra for hcp, 9R, 12R, dhcp and fcc. 
As the Hf sublattice of HfN$_x$ transitions from hcp to 9R to 12R to dhcp and finally to fcc with increasing N concentration, it reduces the number of face-sharing octahedral sites. 
The lowest energy N-vacancy configurations within the hcp, 9R and 12R hosts all avoid the simultaneous occupation of a pair of face-sharing octahedral sites. 
A surprising prediction, however, is the stability of dhcp relative to rocksalt at the HfN composition. 
This is because dhcp contains face sharing octahedral sites that are necessarily occupied in HfN while rocksalt does not. 
As pointed out in Section \ref{sec:zero_kelvin_phase_stability}, the rocksalt form of HfN is thermodynamically favored over the dhcp structure above 670 K due to differences in the vibrational entropy.\cite{zhang2017pressure}
Nevertheless, at low temperatures dhcp HfN is predicted to have a lower energy than the rocksalt structure in spite of the energy cost of having face-sharing nitrogen atoms. 

%While the rocksalt form of HfN is observed experimentally, DFT-PBE predicts that dhcp is the ground state at the stoichiometric Hf:N composition.  \cite{zhang2017pressure}
%This is a surprising prediction since the dhcp crystal structure has face sharing octahedral sites that are all occupied at the HfN composition. 
%As shown in Section \ref{sec:electronic_structure}, pairs of N interstitials in dilute hcp Hf avoid face-sharing octahedral sites due to unfavorable electrostatic interactions between N atoms at short distances. 
%Furthermore, as predicted in Section \ref{sec:zero_kelvin_phase_stability}, an increase in the N concentration results in a change of the stacking sequence of the close-packed Hf layers of the stable HfN$_x$ phases that is accompanied by a reduction in the fraction of face-sharing octahedra. 
%Table \ref{tab:xtal_struct_densities} collects the fraction of face-sharing octahedra for hcp, 9R, 12R, dhcp and fcc. 
%While dhcp has a lower fraction of face-sharing octahedra than hcp and 9R, it has the same fraction as 12R and more than fcc, which has no face-sharing octahedra. 
%When HfN adopts an fcc Hf sublattice, none of the nitrogen interstitials share faces with neighboring nitrogen atoms. 
%Therefore, based solely on the number of face-sharing octahedra, the rocksalt form of HfN should be favored over the dhcp form of HfN. 
%Since dhcp HfN has a lower energy than rocksalt HfN, other factors in addition to the repulsive interaction between face-sharing interstitial nitrogen atoms, play a role in determining the stacking sequence of the Hf sublattice.

\begin{figure}[H]
    \centering
    \includegraphics[width=\linewidth]{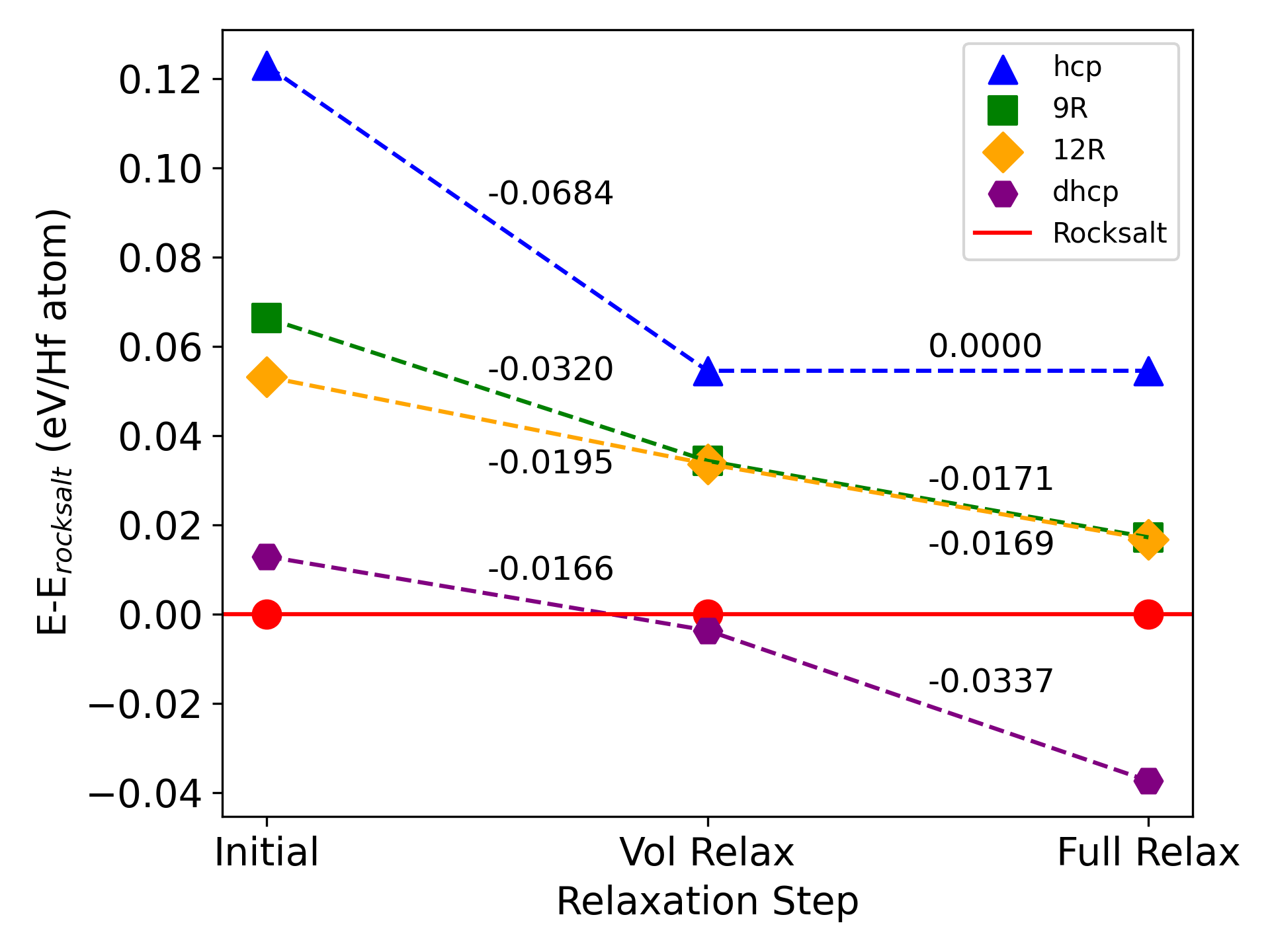}
    \caption{The energy of HfN in the hcp, 9R, 12R, dhcp and rocksalt for different degrees of relaxation as calculated with DFT-PBE. Energies are zeroed to the relaxed rocksalt energy.}
    \label{fig:energy_vs_relaxation_diff_xtal_systems}
\end{figure}

Figure \ref{fig:energy_vs_relaxation_diff_xtal_systems} shows the calculated DFT-PBE energies for stoichiometric HfN having an hcp, 9R, 12R, dhcp and fcc Hf sublattice. 
The first column collects the energies for the ideal, unrelaxed structures, assuming an ideal packing of the Hf atoms and using an identical nearest neighbor Hf-Hf pair distance for each structure. 
The nearest neighbor Hf-Hf distance was set equal to the equilibrium spacing in rocksalt HfN. 
The second column collects the energies after partial relaxation, whereby only the unit cell volume and shape were allowed to relax, but the internal atomic positions were kept at their ideal positions. 
Finally, the third column collects the energies where the lattice vectors and internal atomic positions were allowed to relax. 

In the unrelaxed state, the relative stabilities between the different structures is qualitatively consistent with the fraction of face-sharing octahedra within the structure, with structures having a lower fraction of face-sharing octahedra having a lower energy. 
Nevertheless, there is a large difference in energy between dhcp and 12R even though both structures have the same fraction of face-sharing octahedra. 
Furthermore, the difference in energy between dhcp and 12R is significantly larger than the difference in energy between 12R and 9R, which have a different fraction of face-sharing octahedra. 
Also surprising is that unrelaxed HfN in the dhcp structure has an energy that is very close to that of HfN in the rocksalt structure, which has no face-sharing octahedra. 
When the unit cell volume and shape is allowed to relax, all structures are able to lower their energy relative to that of rocksalt HfN, with the energy of dhcp HfN even dipping below that of the rocksalt structure. 
A further reduction in energy in the 9R, 12R and dhcp structures occurs when internal atomic positions are allowed to relax. 
The dhcp gains more energy during this last relaxation than the 9R and 12R structures because the nitrogen interstitials in dhcp, which occur in pairs, are able to relax further from each other than the nitrogen interstitials of 9R and 12R, which occur in chains of three. 

The favorability of dhcp, even in the unrelaxed state, relative to 12R (which has the same number of face-sharing octahedra) and rocksalt (which has no face-sharing octahedra) likely derives from its more favorable electronic structure.
Figure 7 of the supporting information compares the electronic density of states (DOS) of HfN in the five different crystal structures. 
Only in the dhcp structure does the Fermi level reside in a pseudogap. 
This often occurs when the Fermi level is close to the Brillouin zone boundaries, thereby inducing a lowering (raising) of the occupied (unoccupied) states near the Brillouin boundaries and an enhancement of the cohesive energy of the crystal structure. 
A similar phenomenon occurs in the hcp crystal structure of pure Hf and Zr,\cite{harrison2012electronic}   (Figure 6 in Supporting Information) and is a possible explanation for its stability relative to the fcc and bcc crystal structures.\cite{natarajan2020crystallography} 
The relative stability of different polymorphs of Zr as a function of pressure have also be rationalized by the existence of a pseudogap that separates the occupied electronic states from the unoccupied states.\cite{zhang2010first}
A possible explanation for the stability of dhcp HfN is, therefore, the existence of a pseudogap in the DOS at the Fermi level, with the energetically favorable electronic structure of the host (i.e. a favorable Brillouin zone shape in relation to the Fermi surface) compensating for the repulsive N-N interactions in the face-sharing octahedra of dhcp. 

\begin{table*}
    \centering
    \begin{tabular}{c c c} \hline \hline
         System&  Fraction of face sharing N octahedra & Longest N-N chain\\ \hline 
         hcp&  1& $\inf$\\ 
         9R ($\eta$)&  0.67& 3\\ 
         12R ($\zeta$)&  0.5& 3\\ 
         dhcp&  0.5& 2\\ 
         fcc&  0& N/A\\ \hline \hline 
    \end{tabular}
    \caption{Density of face sharing N octahedra for each crystal system with filled octahedral N interstitials.}
    \label{tab:xtal_struct_densities}
\end{table*}

\subsection{Finite Temperature Phase Diagram}
\label{sec:finite_temperature_phase_diagram}

\begin{figure*}[t!]
    \centering
    \includegraphics[width=\linewidth]{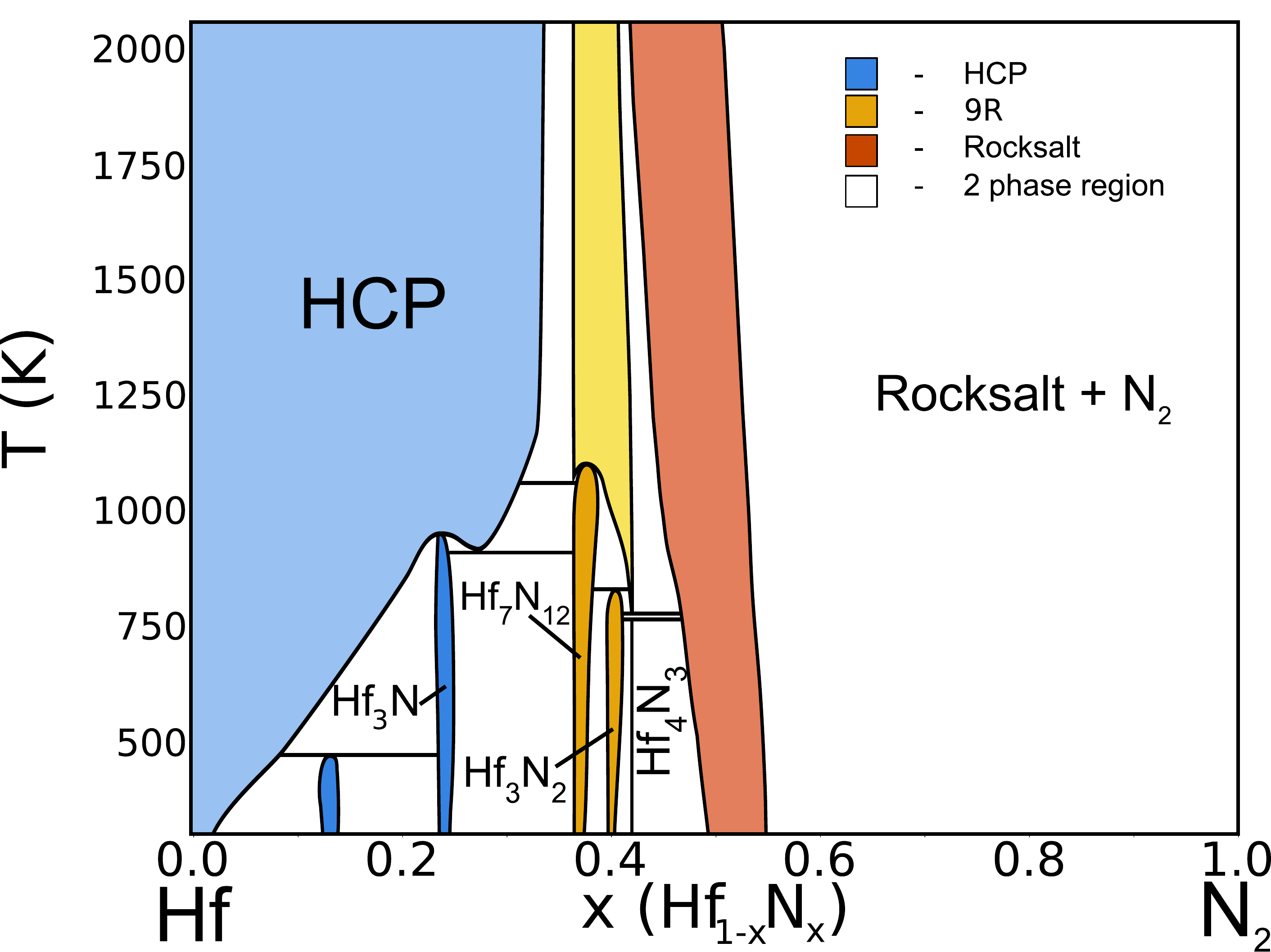}
    \caption{Calculated Hf-N binary phase diagram. Single phase regions with a hcp parent crystal are shown in blue, 9R ($\eta$) regions are shown in gold, and rocksalt phase regions shown in red. Darker colored regions (as depicted in the legend) represent ordered phases in their respective parent crystal structures while lighter colored regions represent disordered phases.}
    \label{fig:hf-n_pd}
\end{figure*}
\begin{figure*}[t!]
    \centering
    \includegraphics[width=\linewidth]{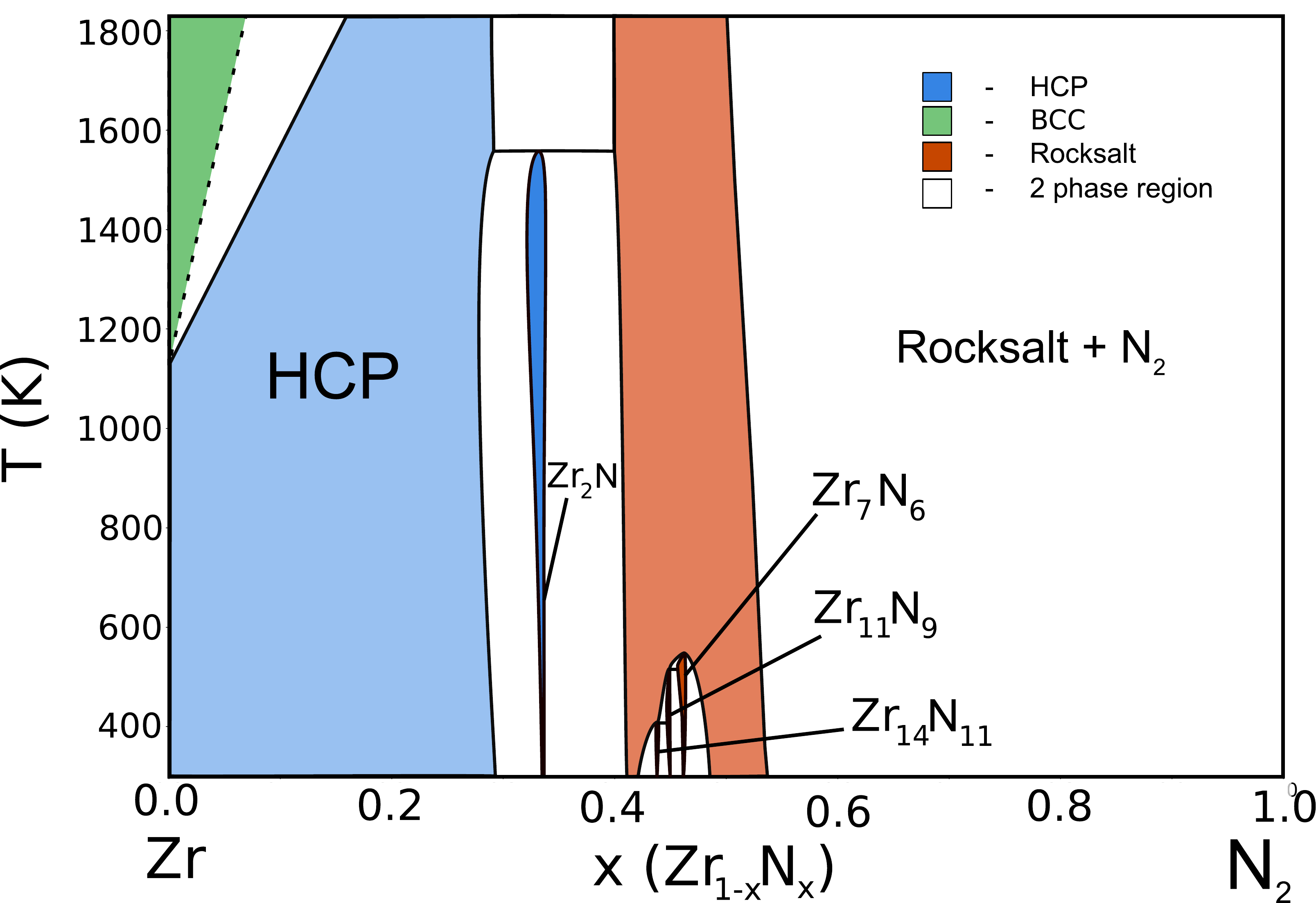}
    \caption{Calculated Zr-N binary phase diagram. Single phase regions with a hcp parent crystal are shown in blue, bcc phase regions are shown in green, and rocksalt phase regions shown in red. Darker colored regions represent ordered phases in their respective parent crystal structures while lighter colored regions represent disordered phases.}
    \label{fig:zr-n_pd}
\end{figure*}

Figure \ref{fig:hf-n_hull} shows that the formation energies of large numbers of nitrogen configurations over the interstitial sites of the Hf hcp, fcc and 9R host structures are close to the convex hull. 
Host structures with low energy configurational excitations relative to their ground state orderings are susceptible to forming solid solutions at elevated temperatures.
To establish the importance of configurational disorder on finite temperature phase stability, we used statistical mechanics methods that rely on cluster expansions to interpolate DFT energies within Monte Carlo simulations.\cite{sanchez1984generalized,de1994cluster,van2018first} 
Cluster expansion Hamiltonians of the configurational energies of the Hf hosts having the hcp, fcc and 9R crystal structures were trained to the large number of DFT-PBE formation energies shown in Figure \ref{fig:hf-n_hull}. 
The cluster expansions, one for each host, were subjected to Monte Carlo simulations to calculate finite temperature thermodynamic properties that enable the construction of free energy models using free energy intergration techniques.\cite{puchala2023casm} 
The $\zeta$-Hf$_4$N$_3$ (12R) ground state was treated as a line compound as the energy penalty to forming anti-site disorder in this structure was found to be significantly higher than in the 9R host.
DHCP is omitted in the calculated phase diagram since it is predicted to become metastable relative to rocksalt above 670 Kelvin when considering vibrational excitations \cite{zhang2017pressure}.
The calculated phase diagrams only account for configurational degrees of freedom, and neglect contributions from vibrational excitations to the free energies of each phase.
Similar cluster expansions were trained for the hcp and fcc host structures of Zr to construct a temperature versus composition phase diagram for the Zr-N binary.

The calculated temperature versus composition phase diagram for the Hf-N binary system is shown in Figure \ref{fig:hf-n_pd}. 
The phases that are derived from the hcp (fcc) host are shown in blue (red), while those that derive from the hybrid stacking fault phases are shown in gold. 
Disordered solid solutions are represented by a lighter blue (gold/red).
The Hf hcp host has multiple ground state nitrogen-vacancy orderings over its interstitial sites at zero kelvin (Figure \ref{fig:hf-n_hull}), but Figure \ref{fig:hf-n_pd} shows that only $\alpha$-Hf$_3$N remains stable above 500 K. % and that it disorders around 1000 K. 
The $\alpha$-Hf$_8$N ground state undergoes a peritectoid transition around 500K while the $\alpha$-Hf$_3$N ground state undergoes an order-disorder transition at 1000K. 
The hcp phase forms a solid solution that is stable as a large single-phase domain, with a nitrogen solubility as high as $x \approx 0.35$ above 1000K. 
The $\beta$-phase (bcc), based on experimental assessments of the Hf-N phase diagram becomes stable above $\sim$ 2016 K.\cite{okamoto1990hf} 
The bcc form of Hf is stabilized by large-scale anharmonic vibrational excitations that require statistical mechanics methods  that go beyond the harmonic approximation to model rigorously.\cite{souvatzis2008entropy,bhattacharya2008mechanical,souvatzis2011temperature,thomas2013finite,kadkhodaei2017free,bechtel2019finite,kadkhodaei2020phonon,jung2023dynamically}

The ordered stacking fault phases $\eta$-Hf$_{12}$N$_7$, $\eta$-Hf$_3$N$_2$, and $\zeta$-Hf$_4$N$_3$ are stable to temperatures between 700 K and 1000 K. 
The $\eta$-Hf$_{3}$N$_2$ and $\eta$-Hf$_{12}$N$_7$ phases accommodate a degree of configurational entropy, which leads to some tolerance for off-stoichiometric compositions at finite temperature. 
%This has not been previously shown in the literature. 
The $\zeta$-Hf$_4$N$_3$ and $\eta$-Hf$_3$N$_2$ phases undergo peritectoid transitions at around 860K and 900K, respectively, while $\eta$-Hf$_{12}$N$_7$ goes through a congruent order-disorder transition around 1200K. 
Above these temperatures, the Hf 9R host forms a disordered solid solution in a narrow composition interval between $x \approx 0.37$ and $x \approx 0.42$, with much of the nitrogen-vacancy disorder occurring over the interstitial sites surrounding the stacking fault. 

The rocksalt phase (having an fcc Hf parent structure) is predicted to be stable above room temperature as a disordered solid solution, shown by the red single phase region in Figure \ref{fig:hf-n_pd}.
The substoichiometric nitrogen-vacancy ordered ground states and the supernitride Hf-vacancy ordered ground states are all predicted to disorder below 300 K. 
The nitrogen solubility in rocksalt Hf$_{1-x}$N$_x$ (right phase boundary of the rocksalt single phase region) is predicted to decrease with increasing temperature. 
This occurs because the sizable configurational entropy due to Hf-vacancy disorder in nitrogen rich rocksalt Hf$_{1-x}$N$_x$ is offset by the stronger temperature dependence of the nitrogen chemical potential in the nitrogen gas due to the translational degrees of freedom of N$_2$ gas molecules. 
The dependence of the nitrogen solubility on adjustments to the nitrogen chemical potential due to uncertainties in the DFT-PBE binding energy of the N$_2$ molecule are discussed further in the Supplementary Information.

The Zr-N phase diagram, shown in Figure \ref{fig:zr-n_pd} only consists of hcp and rocksalt based compounds and solid solutions. 
As is already clear from the zero-Kelvin formation energies in Figure \ref{fig:zr-n hull}, the hybrid stacking-faulted phases are not stable relative to the hcp and fcc derived nitrogen-vacancy ordered phases.
Similar to the Hf-N phase diagram, Zr hcp forms a solid solution at finite temperature with a solubility limit close $x\approx 0.3$, while rocksalt Zr$_{1-x}$N$_x$ forms a wide solid solution between $x \approx 0.4$ and $x \approx 0.53$. 
Four nitrogen-vacancy ordered ground states are predicted to remain stable at temperatures above 300 K: $\alpha$-Zr$_2$N is predicted to transform to an hcp solid solution and a disordered rocksalt at 1560 K through a peritectoid reaction, while the nitrogen-vacancy ordered $\delta$-Zr$_{7}$N$_{6}$, and $\delta$-Zr$_{11}$N$_{9}$ ground states undergo peritectoid reactions at approximately 420 K and 560 K respectively. The $\delta$-Zr$_{14}$N$_{11}$ ground state is predicted to disorder congruently just below 600 K.

\section{Discussion}

The Zr-N and Hf-N binary systems belong to an intriguing class of interstitial alloys.  
Zirconium and hafnium, along with other early transition metals such as Ti, V and Nb, are able to dissolve very high concentrations of nitrogen atoms within octahedrally coordinated interstitial sites of different metallic host crystal structures. %cite -
The resulting solid solutions and compounds exhibit a wide variety of unique structural and functional properties that are of increasing interest for ultra-high temperature structural applications\cite{fahrenholtz2014ultra,fahrenholtz2017ultra} and ultra-low temperature quantum computing devices\cite{cassinese2000transport,potjan2023300}.
In this work, we have performed an in depth first-principles statistical mechanics study of phase stability in the Zr-N and Hf-N binaries, rigorously accounting for the role of configurational disorder on finite temperature phase stability. 

The Zr-N and Hf-N binary systems share many similarities in the types of phases they prefer,\cite{weinberger2017ab,ushakovnavrotskyavdw2019carbidesnitrides} but also exhibit important differences in their temperature versus composition phase diagram. 
Both Zr and Hf favor the hcp crystal structure in pure form up to 1150 K and 2100 K, respectively, above which they transform to the more open bcc crystal structure. 
Zr and Hf can dissolve nitrogen until approximately half of the interstitial octahedral sites of their hcp crystal structures are filled. 
Beyond that composition, nitrogen atoms in hcp are forced to simultaneously occupy face-sharing octahedral sites, which is a highly unfavorable configuration as manifested by a very large nearest neighbor repulsive interaction.
An increase in the nitrogen concentration without filling pairs of face-sharing octahedral sites, therefore, requires a change in the stacking sequence of the close-packed metal layers of the Zr and Hf hcp host structures. 
The ABC stacking of close-packed metal planes in an fcc host is able to accommodate the maximum number of interstitial nitrogen without simultaneously filling pairs of face-sharing octahedral. 
The ZrN and HfN rocksalt phases can be described as consisting of fcc metal sublattices with nitrogen atoms filling all the octahedral interstitial sites. 

At elevated temperatures, both the Zr-N and Hf-N binaries favor hcp based solid-solutions at low to intermediate nitrogen concentrations and rocksalt based compounds at high nitrogen compositions. 
The two systems, however, behave very differently at intermediate nitrogen compositions. 
In the Zr-N binary, the Zr hcp host transitions directly to an fcc host as the composition of nitrogen atoms is increased.
In the Hf-N binary, in contrast, intermediate, stacking-faulted host structures become stable. 
These can be described as consisting of fcc slabs interleaved by stacking faults that locally have an hcp structure.
The existence of these phases was first reported on by Rudy et al \cite{rudy1970crystal}, but it was Weinberger et al \cite{weinberger2019crystal,weinbergerzetaetahfn} who elucidated their crystal structures using first-principles calculations. 
They showed that $\eta$-Hf$_3$N$_2$ and $\zeta$-Hf$_4$N$_3$ have a 9R (ABCBCACAB) metal stacking and a 12R (ABCACABCBCAB) metal stacking, respectively. 
They also discovered $\eta$-Hf$_{12}$N$_7$ as a stable compound, having a 9R Hf metal stacking sequence and a nitrogen-vacancy dumbbell as illustrated in Figure \ref{fig:hf12n7_ordering}. 
In Hf, stacking fault phases become stable since the lattice parameters in the close-packed planes in the fcc and hcp structures are very similar and an interleaving of slabs of fcc with stacking faults having a local hcp environment does not incur any epitaxial strain costs. 
In Zr, this is not the case. 
The hcp and fcc phases have very different lattice parameters within the close-packed planes and a structure that combines fcc and hcp slabs incurs strain energy penalties due to the epitaxial constraints. 

Our calculations of finite temperature phase stability have revealed the importance of configurational disorder in stabilizing not only high temperature solid solutions, but also several substoichiometric Zr and Hf nitrides. 
The importance of configurational entropy within the stacking faulted phases has not been previously shown.
In fact, the low temperature nitrogen-vacancy orderings over the interstitial sites of Hf 9R disorder to form a stable 9R solid solution above approximately 1000 K. 
The rocksalt ZrN and HfN phases can also tolerate a high degree of off stoichiometry through the introduction of nitrogen vacancies to make subnitrides or metal vacancies to make supernitrides.
The vacancies remain disordered to temperatures as low as 500 K. 
The HfN and ZrN supernitrides are stable relative to N$_2$ gas up to approximately 1900 K and 1700 K, respectively.
The phase boundary for rocksalt to N$_2$ is fairly robust to shifts in the nitrogen chemical potential, as explained in more detail in the Supporting Information.
Of note, our study neglects the effects of vibrations, so the dhcp-rocksalt transition is not modeled in our finite temperature phase diagram.
Additionally, it does not capture the hcp to bcc transition in metallic Hf/Zr that is stabilized due to vibrations at high temperature.
While the dhcp crystal structures of HfN and ZrN have unfavorable N-N interactions due to the simultaneous occupancy of face-sharing octahedral sites by nitrogen interstitials, they are nevertheless predicted to have a lower energy than the rocksalt crystal structures of HfN and ZrN.
A clear explanation for this prediction is not directly evident, however, it likely has its origin in key differences in the electronic structure between the two crystals, with the Fermi levels of the dhcp HfN and ZrN compounds coinciding with a pseudo gap in their electronic density of states.

While this study focused on the thermodynamic properties of Zr-N and Hf-N phases, very little remains known about the kinetic properties of transition metal nitrides.
%metal and nitrogen diffusion coefficients in transition metal nitrides. 
The non-dilute concentrations of interstitial nitrogen are likely to lead to highly correlated diffusion mechanisms. 
Interstitial diffusion within close-packed frameworks is known to be mediated by vacancy clusters (i.e. divacancies and triple vacancies) as shown in other rocksalt and close-packed layered compounds. \cite{van2008nondilute,bhattacharya2011first,van2013understanding,kolli2021elucidating}
The mobililty of interstitial nitrogen atoms is especially important within the fcc metal frameworks of the rocksalt phases in high temperature applications, where nitrogen off-gassing is more likely. 
Also of interest is to establish whether the metal atoms themselves are mobile within the rocksalt phases. 
Another intriguing aspect about the kinetic properties of early transition metal nitrides is the crystallographic mechanism with which the changes in stacking sequence are realized upon the addition of nitrogen to the host. 
One mechanism could rely on the passage of partial dislocations to mediate a change in stacking sequence, starting from the hcp crystal structure of the pure metal going to the fcc structure of the final rocksalt phases. 
This mechanism would proceed through the passage of semi-coherent interfaces, made up of an array of partial dislocations that migrate under the chemical driving force exerted by an influx of nitrogen atoms. 
Similar phenomena occur in layered intercalation compounds used in Li-ion and Na-ion batteries.\cite{gabrisch2002character,kaufman2019understanding}
Changes in the metal stacking sequence can also occur by a reconstructive mechanism, with as consequence that the nitride rich phases have very limited orientational relationships with the original metal structures. 
The mechanisms and rates of nitrogen diffusion and phase transformations between the various stable and metastable nitrides will have important consequences for manufacturing approaches of structures and devices that rely on specific Hf-N or Zr-N phases. 

%Structural transformations between different stacking sequences - passage of partial dislocations or by means of BCC and a combination of Bain and Burgers \cite{raju2019toward}

\section{Conclusion}
We have performed a comprehensive first principles study of phase stability in the Hf-N and Zr-N binary systems. 
In addition to verifying stability of previously computed and experimentally observed phases, we discovered new ordered phases at both 0K and at finite temperature. % and have analytically accounted for the N$_2$ chemical potential with respect to temperature. 
This includes stable ordered phases in hcp Hf-N and Zr-N, and order-disorder transitions in the stacking fault phases in Hf-N. 
Additionally, we found that the rocksalt phase does not allow for simultaneous Zr/Hf and N vacancies. %, contrary to experimental predictions from over 50 years ago. 
We determined that stacking fault phase stability in Hf-N but not Zr-N is due to in-plane lattice mismatch between hcp and fcc based end members. 
The complete explanation for why DFT predicts DHCP HfN to be significantly lower energy than rocksalt HfN remains undetermined and is grounds for promising future work.

\section{Data Availability}
All data will be made available through Materials Commons \cite{puchala2016materials}.

\subsection{Acknowledgements}
%This work was made possible with the support of the ONR BRC Program, Grant Number N00014-18-1-2392.
This work was supported by the U.S. Department of Energy, Office of Basic Energy Sciences, Division of Materials Sciences and Engineering under Award \#DE-SC0008637 as part of the Center for
Predictive Integrated Structural Materials Science (PRISMS Center)
This work was made possible using of computational facilities purchased with funds from the National Science Foundation (CNS-1725797) and administered by the Center for Scientific Computing (CSC). The CSC is supported by the California NanoSystems Institute and the Materials Research Science and Engineering Center (MRSEC; NSF DMR 2308708) at UC Santa Barbara.
The research reported here made use of the shared facilities of the Materials Research Science and Engineering Center (MRSEC) at UC Santa Barbara: NSF DMR–2308708. The UC Santa Barbara MRSEC is a member of the Materials Research Facilities Network (www.mrfn.org).
This research used resources of the National Energy Research Scientific Computing Center (NERSC), a Department of Energy Office of Science User Facility using NERSC award BES-ERCAP0026626.

\bibliography{./references.bib}

\end{document}